\documentstyle[epsf]{article} 
\begin{document}
\title{Generic wormhole throats}
\author{
Matt Visser$^{\dag}$ and David Hochberg$^{\ddag}$\\
\\
$^{\dag}$Physics Department, Washington University\\
Saint Louis, Missouri 63130-4899, USA\\
\\
$^{\ddag}$Laboratorio de Astrof\'{\i}sica Espacial y F\'{\i}sica Fundamental\\
Apartado 50727, 28080 Madrid, Spain\\
}
\date{15 October 1997}
\maketitle
\centerline{$^{\dag}$ e-mail: visser@kiwi.wustl.edu}

\centerline{$^{\ddag}$ e-mail: hochberg@laeff.esa.es}

\begin{center}
gr-qc/9710001\\
To appear in the proceedings of the Haifa Workshop: \\
The Internal Structure of Black Holes\\
and Spacetime Singularities.
\end{center}
\bigskip
\begin{abstract}
Wormholes and black holes have, apart from a few historical oddities,
traditionally been treated a quite separate objects with relatively
little overlap. The possibility of a connection arises in that
wormholes, if they exist, might have profound influence on black
holes, their event horizons, and their internal structure. For
instance: (1) small wormhole-induced perturbations in the geometry
can lead to massive non-perturbative shifts in the event horizon,
(2) Planck-scale wormholes near any spacelike singularity might
let information travel in effectively spacelike directions, (3)
vacuum polarization effects near any singularity might conceivably
lead to a ``punch through'' into another asymptotically flat
region---effectively transforming a black hole into a wormhole.
After discussing these connections between black hole physics and
wormhole physics we embark on an overview of what can generally be
said about traversable wormholes and their throats. We discuss the
violations of the energy conditions that typically occur at and
near the throat of any traversable wormhole and emphasize the
generic nature of this result. We discuss the original Morris--Thorne
wormhole and its generalization to a spherically symmetric time-dependent
wormhole. We discuss spherically symmetric Brans--Dicke wormholes
as examples of how to hide the energy condition violations in an
inappropriate choice of definitions. We also discuss the relationship
of these results to the topological censorship theorem.  Finally
we turn to a rather general class of wormholes that permit explicit
analysis: generic static traversable wormholes (without any symmetry).
We show that topology is too limited a tool to accurately characterize
a generic traversable wormhole---in general one needs geometric
information to detect the presence of a wormhole, or more precisely
to locate the wormhole throat. For an arbitrary static spacetime
we shall define the wormhole throat in terms of a $2$--dimensional
constant-time hypersurface of minimal area.  (Zero trace for the
extrinsic curvature plus a ``flare--out'' condition.) This enables
us to severely constrain the geometry of spacetime at the wormhole
throat and to derive generalized theorems regarding violations of
the energy conditions---theorems that do not involve geodesic
averaging but nevertheless apply to situations much more general
than the spherically symmetric Morris--Thorne traversable wormhole.
[For example: the null energy condition (NEC), when suitably weighted
and integrated over the wormhole throat, must be violated.]
\end{abstract}


\section{Introduction}
\def\tr{\hbox{\rm tr}}
\def\implies{\Rightarrow}
\def\conv{\hbox{\rm conv}}
\def\Re{ {\cal R} }

Traversable wormholes~\cite{Morris-Thorne,MTY,Visser} have
traditionally been viewed as quite distinct from black holes, with
essentially zero overlap in techniques and topics. (Historical
oddities that might at first seem to be counter-examples to the
above are the Schwarzschild wormhole, which is not traversable,
and the Einstein--Rosen bridge, which is simply a bad choice of
coordinates on Schwarzschild spacetime~\cite[pages 45--51]{Visser}.)
Since this workshop is primarily directed toward the study of black
holes we shall start by indicating some aspects of commonality and
inter-linkage between these objects.

\begin{figure}[htb]
\vbox{\hfil\epsfbox{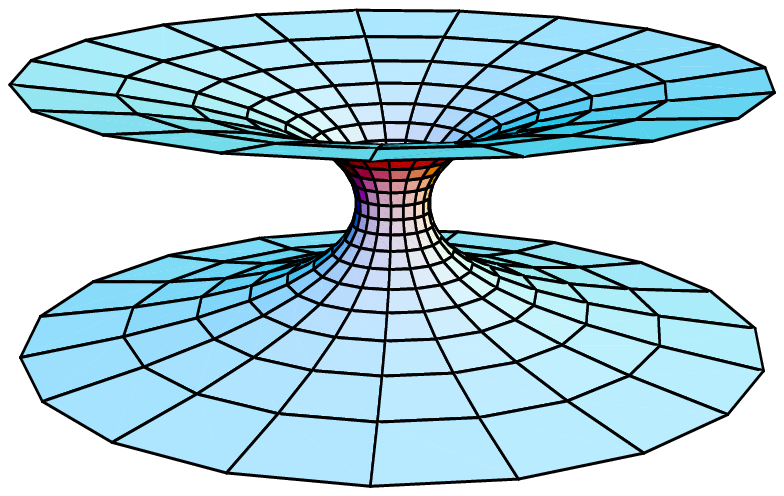}\hfil}
\caption
{\label{F-inter-universe}
Schematic representation of an inter-universe wormhole connecting
two asymptotically flat regions.}
\end{figure}

First: We point out that traversable wormholes, if they exist, can
have violent non-perturbative influence on black hole event horizons.
This is fundamentally due to the fact that the event horizon is
defined in a global manner, so that small changes in the geometry,
because they have all the time in the universe to propagate to
future null infinity, can have large effects on the event horizon.

Second: We point out that the region near any curvature singularity is
expected to be subject to large fluctuations in the metric. If
Wheeler's spacetime foam picture is to be believed, the region near a
curvature singularity should be infested with wormholes. If it's a
spacelike singularity (meaning, it's spacelike before you take the
wormholes into account) then the wormholes can connect regions that
would otherwise be outside each other's lightcones, and so lead to
information transport in an effectively spacelike direction. You still
have to worry about how to transfer the information across the event
horizon, but this is a start towards a lazy way out of the black hole
information paradox. (The information paradox is only a paradox if you
take the Schwarzschild singularity too seriously, in particular if you
take the spacelike nature of the singularity too seriously.)

Third: In addition to the fluctuations that take place near a
curvature singularity, we would also expect large effects on the
expectation value of the metric due to gravitational vacuum
polarization. Since gravitational vacuum polarization typically
violates the energy conditions, this could lead us to expect a
``punch through'' to another asymptotically flat region. The
possibility that singularities might heal themselves by automatic
conversion into wormhole throats, though maybe unlikely, should at
least be kept in mind.

After developing these linkages between black hole physics and
wormhole physics, we embark on an overview of traversable wormholes,
concentrating on the region near the throat of the wormhole.  One
of the most important features that characterizes traversable
wormholes is the violation of the energy conditions, in particular
the null energy condition, {\em at or near the throat}. The energy
condition violations were first discovered in the static spherically
symmetric Morris--Thorne wormholes, but the result is generic
(modulo certain technical assumptions) as borne out by the topological
censorship theorem, and also by the generic analysis of static
wormholes developed later in this survey.

We shall discuss the case of time-dependent spherically symmetric
wormholes, wherein the energy violation conditions can be isolated at
particular regions in time (in the same way that static thin-shell
wormholes permit one to isolate the energy condition violations at
particular regions of space.) We show that these results are
compatible with the topological censorship theorem.  Perhaps
surprisingly, we show that cosmological inflation is useless in terms
of generating even temporary suspension of the energy condition
violations.

We further discuss the case of spherically symmetric Brans--Dicke
wormholes as an example of what happens in non--Einstein theories of
gravity: In this case it is possible to mistakenly conclude that the
energy conditions are not violated, but this would merely be a
consequence of an inappropriate choice of definitions. For instance:
If one works in the Einstein frame, defines the existence of a
wormhole in terms of the Einstein metric, and calculates using the
total stress energy tensor, then the null energy condition must be
violated at or near the throat. If one artificially divides the
total stress-energy into (Brans--Dicke stress-energy) plus (ordinary
stress-energy), then since the Brans--Dicke stress energy (calculated
in the Einstein frame) sometimes violates the energy conditions (if the
Brans--Dicke parameter $\omega$ is less than $-3/2$), the ``ordinary'' part
stress-energy can sometimes satisfy the null energy condition.

Similarly, suppose one works in the Jordan frame, and defines the
existence of a wormhole using the Jordan metric. If one calculates the
total stress energy tensor then it is still true that the null energy
condition must be violated at or near the throat.  If one now
artificially divides the total stress-energy into (Brans--Dicke
stress-energy) plus (ordinary stress-energy), then for suitable
choices of the Brans--Dicke parameter $\omega$ ($\omega<-2$) one can
hide all the energy condition violations in the Brans-Dicke field and
permit the ordinary stress-energy to satisfy the energy conditions.

To further confuse the issue, one could define the existence (or
nonexistence) of a wormhole using one frame, and then calculate the
Einstein tensor and total stress-energy in the other frame.  This is a
dangerous and misleading procedure: We shall show that the definition
of the existence and location of a wormhole throat is not frame
independent (because it is not conformally invariant).  Jumping from
one frame to the other in the middle of the calculation can easily
lead to meaningless results.  It is critical to realise that the
energy condition violations must still be there and in fact are still
there: they have merely been hidden by sleight of hand.
Qualitatively, these comments also apply to other non--Einstein
theories of gravity such as the Einstein--Cartan theory, Dilaton
gravity, Lovelock gravity, Gauss--Bonnet gravity, etc...

Finally, we wrap up by presenting a general analysis for static
traversable wormholes that completely avoids all symmetry requirements
and even avoids the need to assume the existence of asymptotically
flat regions.  This exercise is particularly useful in that it
places the notion of wormhole in a much more general setting.
Indeed, wormholes are often viewed as intrinsically topological
objects, occurring in multiply connected spacetimes. The Morris--Thorne
class of inter-universe traversable wormholes is even more restricted,
requiring both exact spherical symmetry and the existence of two
asymptotically flat regions in the spacetime.  To deal with
intra-universe traversable wormholes, the Morris--Thorne analysis
must be subjected to an approximation procedure wherein the two
ends of the wormhole are distorted and forced to reside in the same
asymptotically flat region. The existence of one or more asymptotically
flat regions is an essential ingredient of the Morris--Thorne
approach~\cite{Morris-Thorne}.

\begin{figure}[htb]
\vbox{\hfil\epsfbox{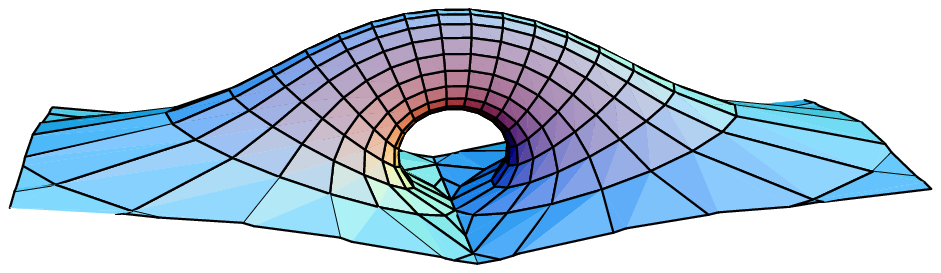}\hfil}
\caption
{\label{F-intra-universe}
Schematic representation of an intra-universe wormhole formed by
deforming an inter-universe wormhole and forcing the two asymptotically
flat regions to merge.}
\end{figure}

However, there are many other classes of geometries that one might
still quite reasonably want to classify as wormholes, that either
have trivial topology~\cite{Visser}, or do not possess any
asymptotically flat region~\cite{HPS}, or exhibit both these
phenomena.

\begin{figure}[htb]
\vbox{\hfil\epsfbox{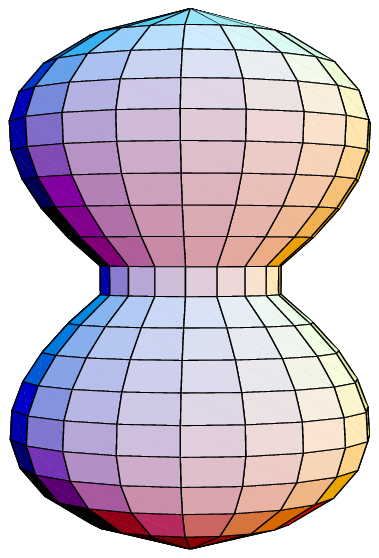}\hfil}
\caption[Dumbbell wormhole]
{\label{F-FRW}
A ``dumbbell wormhole'': Formed (for example) by two closed
Friedman--Robertson--Walker spacetimes connected by a narrow neck.}
\end{figure}

\begin{figure}[htb]
\vbox{\hfil\epsfbox{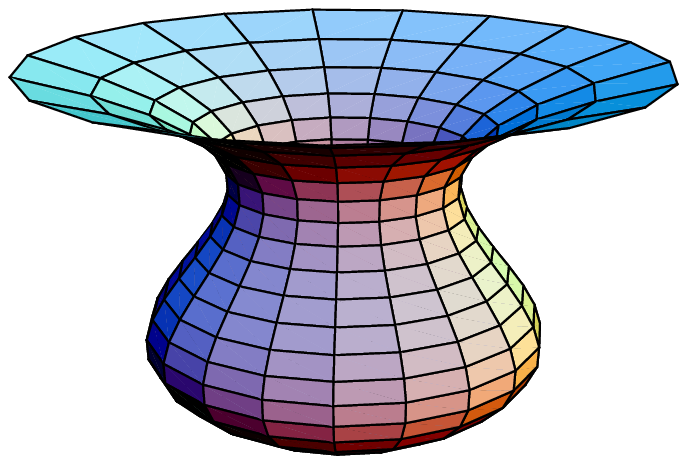}\hfil}
\caption[Topologically trivial wormhole]
{\label{F-trivial}
A wormhole with trivial topology: Formed (for example) by connecting
a single closed Friedman--Robertson--Walker spacetime to Minkowski
space by a narrow neck.}
\end{figure}

A simple example of a wormhole lacking an asymptotically flat region
is two closed Friedman--Robertson--Walker spacetimes connected by
a narrow neck (see figure~\ref{F-FRW}), you might want to call this
a ``dumbbell wormhole''. A simple example of a wormhole with trivial
topology is a single closed Friedman--Robertson--Walker spacetime
connected by a narrow neck to ordinary Minkowski space (see
figure~\ref{F-trivial}).  A general taxonomy of wormhole exemplars
may be found in~\cite[pages 89--93]{Visser}, and discussions of
wormholes with trivial topology may also be found in~\cite[pages
53--74]{Visser}.

To set up the analysis for a generic static throat, we first have
to define exactly what we mean by a wormhole---we find that there
is a nice {\em geometrical\,} (not topological) characterization
of the existence of, and location of, a wormhole ``throat''. This
characterization is developed in terms of a hypersurface of minimal
area, subject to a ``flare--out'' condition that generalizes that
of the Morris--Thorne analysis.

With this definition in place, we can develop a number of theorems
about the existence of ``exotic matter'' at the wormhole throat.
These theorems generalize the original Morris--Thorne result by
showing that the null energy condition (NEC) is generically violated
at some points on or near the two-dimensional surface comprising
the wormhole throat.  These results should be viewed as complementary
to the topological censorship theorem~\cite{Topological-censorship}.
The topological censorship theorem tells us that in a spacetime
containing a traversable wormhole the averaged null energy condition
must be violated along at least some (not all) null geodesics, but
the theorem provides very limited information on where these
violations occur. The analysis of this paper shows that some of
these violations of the energy conditions are concentrated in the
expected place:  on or near the throat of the wormhole. The present
analysis, because it is purely local, also does not need the many
technical assumptions about asymptotic flatness, future and past
null infinities, and global hyperbolicity that are needed as
ingredients for the topological censorship
theorem~\cite{Topological-censorship}.  

The key simplifying assumption in the present analysis is that of
taking a static wormhole. While we believe that a generalization
to dynamic wormholes is possible, the situation becomes technically
much more complex and and one is rapidly lost in an impenetrable
thicket of definitional subtleties and formalism. (A suggestion,
due to Page, whereby a wormhole throat is viewed as an {\em
anti-trapped surface} in spacetime holds promise for suitable
generalization to the fully dynamic case~\cite{Page}.)

In summary, the violations of the energy conditions at wormhole
throats are unavoidable. Many of the of the attempts made at
building a wormhole without violating the energy conditions do so
only by hiding the energy condition violations in some subsidiary
field, or by hiding the violations at late or early times.

\section{Connections}

We start by describing a few scenarios whereby wormholes might prove
of interest to black hole physics. Don't take any of these scenarios
too seriously: they are presented more as suggestions for things
to consider than as definite proposals for serious models.

\subsection{Non-perturbative changes in the event horizon}

Traversable wormholes, if they exist, can lead to massive
nonperturbative changes in the event horizon of a black hole at
the cost of relatively minor perturbations in the geometry. (This
should not be too surprising: it is a general feature of Einstein
gravity that small perturbations in the geometry can lead to massive
perturbations in the event horizon.) One of the simplest examples
of this effect is to take a wormhole (with two mouths), and a black
hole, and then throw one wormhole mouth down into the black hole
while keeping the second wormhole mouth
outside~\cite{Frolov-Novikov:device}. Everything in the past
lightcone of the mouth that falls into the wormhole, including the
segment that is behind where the event horizon used to be before
the wormhole mouth fell in, can now influence future null infinity
by taking a shortcut through the wormhole.

This can best be seen visually; we idealize the wormhole to have
a mouth extremely small radius, and model the wormhole by a pair
of timelike lines in spacetime that are mathematically identified.
The black hole will be idealized in the usual way with a Penrose
diagram.  The geometry of the spacetime is affected by the wormhole
only in the immediate vicinity of the wormhole mouths, but the
global properties (such as the event horizon) suffer drastic
non-perturbative changes.

\begin{figure}[htb]
\vbox{\hfil\epsfbox{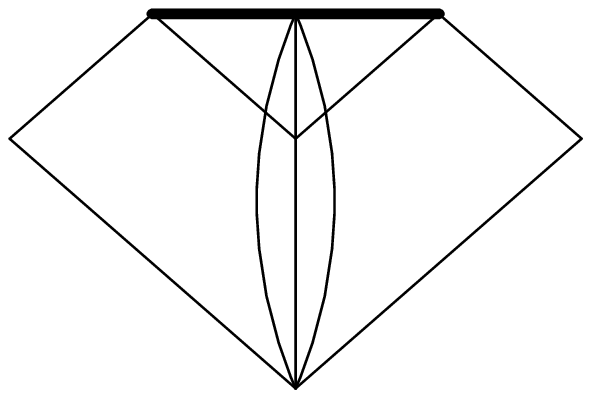}\hfil}
\caption
{\label{F-penrose-astro}
Penrose diagram for an astrophysical black hole formed by (say)
stellar collapse.}
\end{figure}

\def\dead-code{
\begin{figure}[htb]
\vbox{\hfil\epsfbox{F-astro-plus-wormhole.eps}\hfil}
\caption
{\label{F-astro-plus-wormhole}
Penrose diagram for an astrophysical black hole that has swallowed
one mouth of a traversable wormhole. Note the violent alteration
in the location of the event horizon, although the changes in the
geometry are confined to the regions immediately near the wormhole
mouths.}
\end{figure}
}

With a swarm of traversable wormholes one can chip away at the event
horizon to the extent that it ``almost'' disappears. The event horizon
is progressively eaten away as more wormhole mouths fall into the
region behind the apparent horizon.

\def\dead-code{
\begin{figure}[htb]
\vbox{\hfil\epsfbox{F-astro-chip.eps}\hfil}
\caption[Penrose diagram for a black hole]
{\label{F-astro-chip}
Penrose diagram for an astrophysical black hole that has swallowed
a swarm of traversable wormhole mouths. Assume that for each
traversable wormhole only one mouth falls in, the other mouth
remaining outside.}
\end{figure}
}

\subsection{Effectively spacelike information transfer}

Near any spacetime curvature singularity, in the region where
curvatures are large, the spacetime foam picture developed by
Wheeler~\cite{Wheeler:56,Wheeler:57} strongly suggests that Lorentzian
wormholes will be rapidly popping in and out of existence. (We
ignore for the sake of present discussion potential difficulties
associated with topology change in Lorentzian manifolds: certainly
something peculiar has to happen in the region where the wormhole
is created or destroyed. See~\cite[pages 61--73]{Visser}.)

We can think of at least two plausible models for Lorentzian wormhole
creation. (We again take the approximation that the wormholes have
infinitely small mouth radius, and so can be modelled by infinitely
thin spacetime world-lines.) In model (1)  the two mouths are created
(pulled out of the Planck slop?) at different places, and after
creation one simply has two timelike world-lines that are to be
identified. In model (2) the two mouths are created at the same
spacetime point and the two mouths then subsequently move off along
distinct timelike world-lines (which are again identified to produce
the wormhole structure).

In model (1) the spacetime distribution of the creation points for
the two mouths is something we have no idea how to calculate. we
should hope that the two creation points are spacelike separated
since otherwise the Chronology Protection Conjecture~\cite{Hawking:cpc}
has failed at the outset. Beyond that, very little can be said. In
model (2) the Chronology Protection Conjecture is respected at the
initial creation point, but one still has to worry about the
subsequent evolution of the wormhole mouths. Since all of this is
presumably taking place deep inside the Planck slop, issues of
reliability of the entire semiclassical approximation also deserve
attention~\cite{Visser:reliability}.

\def\dead-code{
\begin{figure}[htb]
\vbox{\hfil\epsfbox{F-create-1.eps}\hfil}
\caption
{\label{F-create-1}
Model 1: Penrose diagram for a wormhole being ripped out of the
Planck slop near a spacelike singularity. The two mouths are created
at points that are spacelike separated and each mouth subsequently
follows a timelike world-line before colliding with the singularity.}
\end{figure}
} 

\def\dead-code{
\begin{figure}[htb]
\vbox{\hfil\epsfbox{F-create-2.eps}\hfil}
\caption
{\label{F-create-2}
Model 2: Penrose diagram for a wormhole being ripped out of the
Planck slop near a spacelike singularity.  The two mouths are
created at the same spacetime pint, separate,  and each mouth
subsequently follows a timelike world-line before colliding with
the singularity.}
\end{figure}
} 

Whichever of these models (1) or (2) you pick [and both models are
perfectly compatible with both the typical noises generated in this
field, and our current ignorance of quantum gravity] the wormhole
mechanism takes some points that would be spacelike separated if
the wormholes were not present and makes them timelike separated.
This is quite sufficient to allow ``effectively spacelike'' information
transfer in a thin region near the curvature singularity.

\def\dead-code{
\begin{figure}[htb]
\vbox{\hfil\epsfbox{F-spacelike.eps}\hfil}
\caption
{\label{F-spacelike}
``Effectively spacelike'' information transfer near a spacelike
singularity.  The wormhole gas takes points that would be spacelike
separated in the absence of the wormholes and connects them in a
timelike manner. Information can then flow in a ``spacelike''
direction along the spacelike singularity.}
\end{figure}
}

To actually get information out of the black hole, you will either
have to live with an extension of model (1) wherein the second
wormhole mouth pops into existence just outside the event horizon,
or rely on an external wormhole pair one of whose mouths is permitted
to fall into the horizon.

\subsection{``Punch-through'' near the singularity}

Near a spacelike curvature singularity, the quantum vacuum expectation
value of the stress-energy tensor is likely to be heading off to
infinity due to gravitational vacuum polarization effects. But
gravitational vacuum polarization quite typically leads to violations
of the energy conditions~\cite{gvp:anec,gvp1,gvp2,gvp3,gvp4}. And
violations of the energy conditions are a generic feature of
wormholes.

This suggests (hints) that curvature singularities might heal
themselves by ``punching through'' to another asymptotically flat
region in a manner similar to a wormhole. (See for example the
minisuperspace model discussed in~\cite[pages 347--359]{Visser}
and~\cite{Visser:quantum-wormholes,Visser:wormshop,Visser:rice}.)
For a concrete suggestion along these lines consider the metric

\begin{eqnarray}
ds^2 &=& 
- \left(1 - {2GM\over\sqrt{\ell^2+r^2}}\right) dt^2 
+ \left(1 - {2GM\over\sqrt{\ell^2+r^2}}\right)^{-1} dr^2 
\nonumber\\
&&
+ (\ell^2+r^2) \left\{ d\theta^2 + \sin^2\theta \; d\phi^2 \right\}.
\end{eqnarray}
This is almost the Schwarzschild geometry, apart from the parameter
$\ell$. The radial variable $r$ can now be extended all the way from
$+\infty$ to $-\infty$, and there is a symmetry under interchange
$r\to -r$. For $\ell \ll 2GM$ (but $\ell\neq0$) and $r$ not too close
to zero (which is where the spacelike singularity is for $\ell=0$),
this is indistinguishable from the Schwarzschild solution.  There is a
thin region near $r=0$ where the stress-energy is appreciably
different from zero. In fact the Einstein tensor is

\begin{eqnarray}
G_{\hat t\hat t} 
&=&
- {\ell^2 \left(1-{4GM\over\sqrt{\ell^2+r^2}}\right)
  \over
  (\ell^2+r^2)^2},
\\
G_{\hat\theta \hat\theta} 
&=&
+ {\ell^2 \left(1-{GM\over\sqrt{\ell^2+r^2}}\right)
  \over
  (\ell^2+r^2)^2},
\\
G_{\hat r \hat r} 
&=&
- {\ell^2\over (\ell^2+r^2)^2},
\end{eqnarray}
Note that this is not a traversable wormhole in the usual sense,
since the region near the ``throat'' ($r=0$) it is the the radial
direction that is timelike. Thus travel is definitely one-way, and
if anything one should call this a ``timehole''.  The hope of course
is that some sort of geometry similar to the above might emerge as
a self-consistent solution to the semiclassical field equations.
A timelike version of the Hochberg--Popov--Sushkov analysis~\cite{HPS}
is what we have in mind. Note in particular that at $r=0$ we have

\begin{equation}
G_{\hat t\hat t} + G_{\hat r \hat r} 
=
+ {(4GM-2\ell) \over \ell^3}.
\end{equation}
Thus the null energy condition is {\em not} violated at $r=0$,
indicating that one-way ``timeholes'' of this type are qualitatively
different from the two-way traversable wormholes considered in the
rest of this survey. (There are some similarities here with the
Aichelburg--Schein wormholes, which are not true traversable wormholes
in that they are only one-way traversable~\cite{Aichelburg}. In
contrast the Aichelburg--Israel--Schein wormholes are true traversable
wormholes which violate the energy conditions in the usual
manner~\cite{Aichelburg2}.)

\begin{figure}[htb]
\vbox{\hfil\epsfbox{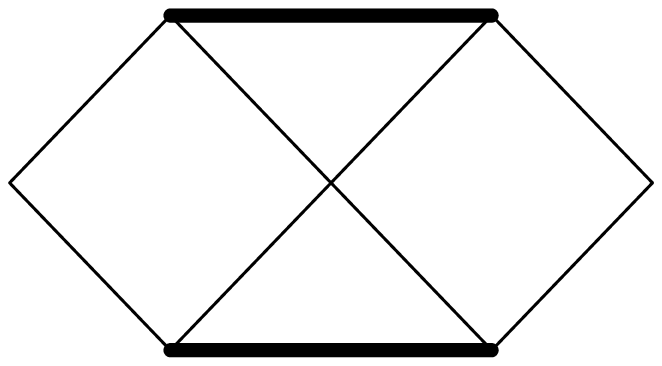}\hfil}
\caption
{\label{F-maximal}
Penrose diagram for the maximally extended Schwarzschild geometry.}
\end{figure}

\begin{figure}[htb]
\vbox{\hfil\epsfbox{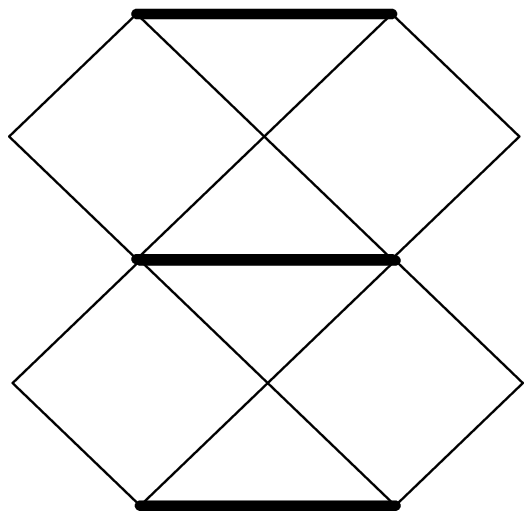}\hfil}
\caption
{\label{F-timehole}
Penrose diagram for a ``timehole''. Gravitational vacuum polarization
near where the spacelike singularity would have been could plausibly
lead to ``punch through'' to another asymptotically flat region.}
\end{figure}

\subsection{``Punch-through'' near the horizon}

A bolder proposal is that ``punch through'' might occur not near where
the central singularity would have been, but instead that ``punch
through'' might occur out where one would naively expect the event
horizon to be.  For instance, if one insists on using the Boulware
vacuum, and insists on a spacetime containing an event horizon, then
the renormalized stress-energy will diverge at the horizon. Therefore,
if one insists on having a self-consistent solution to the
semiclassical field equations in the Boulware vacuum, then no event
horizon can survive. Given that test-field calculations on
Schwarzschild spacetime show infinite gravitational vacuum
polarization at the would-be event horizon, and that this
gravitational vacuum polarization violates the null energy condition,
one might suspect that the self-consistent solution contains a
traversable wormhole.

Bolstering this suspicion is the fact that the Hochberg--Popov--Sushkov
analysis~\cite{HPS} has numerically integrated the fourth-order
differential equations derived from semiclassical gravity, and
found wormhole throats in the Boulware vacuum. (There are a tangle
of technical issues here to do with the size of the wormhole throat
and whether or not it is possible to extend these numerical solutions
out into truly asymptotically flat regions. We refer the interested
reader to the literature.)

A simple and explicit toy model geometry along these lines is to
start with the Schwarzschild solution written in isotropic coordinates

\begin{eqnarray}
ds^2 &=& 
- \left({1-GM/2r\over1+GM/2r}\right)^2 dt^2 
\nonumber\\
&&
+ (1+GM/2r)^4 \left\{ dr^2 + r^2 [ d\theta^2 + \sin^2 d\phi^2 ] \right\}.
\end{eqnarray}
The radial coordinate $r$ runs from $r=0$ to $r=+\infty$. The horizon
is at $r=R\equiv GM/2$, and there is symmetry under radial inversion
$r\to R^2/r$.  Now perturb this geometry by distorting the $g_{tt}$
component of the metric:

\begin{eqnarray}
ds^2 &=& 
-\left\{
{(R^2-r^2)^2 + \ell^2 r^2 \over (R+r)^4+ \ell^2 r^2 }
\right\} dt^2
\nonumber
\\
&&
+ (1+R/r)^4 \left\{ dr^2 + r^2 [ d\theta^2 + \sin^2 d\phi^2 ] \right\}.
\end{eqnarray}
The radial coordinate $r$ still runs from $r=0$ to $r=+\infty$ and
there is still a symmetry under radial inversion $r\to R^2/r$. The
particular form of the distortion given above has been chosen so
that the asymptotic form of the metric at spatial infinity agrees
with the Schwarzschild metric up to order $O[r^{-3}]$.  But $r = R
\equiv GM/2$ is no longer an event horizon, it is now the location of
a wormhole throat connecting two asymptotically flat regions (one
at $r=+\infty$ and the other at $r=0$). The throat is at finite
redshift $1+z = \sqrt{1+(16 R^2/\ell^2)}$.  For $\ell \ll R$ the
region outside the would-be event horizon is arbitrarily close to
Schwarzschild geometry. (This is an example of a ``proximal
Schwarzschild'' wormhole as discussed in~\cite[pages 147--149]{Visser}.)
The stress-energy tensor is a bit messy, but falls of as order
$O[r^{-5}]$ for large $r$.  Thus there is a thin layer of energy
condition violating matter near the throat of the wormhole, but by
the time you are a little distance away from the throat you cannot
tell the difference between this geometry and the usual Schwarzschild
solution.

\subsection{Summary}

The key point to be extracted from the above discussion is that
small perturbations on the geometry of spacetime can make drastic
perturbations to the notion of a black hole. Small perturbations
of the geometry can eat away at the event horizon like cancer, can
drastically affect the properties of spacelike singularities
(spacelike information transfer, time holes), and in the wrong
hands can even make event horizons vanish (proximal Schwarzschild
wormholes). All of the specific examples discussed above use
wormholes, or things that are almost wormholes, so we feel it is
a good idea for black hole physicists to have some basic understanding
of the wormhole system and its limitations.


\section{Energy conditions: An overview}

The Morris--Thorne analysis~\cite{Morris-Thorne} revitalized interest
in Lorentzian traversable wormholes. Morris and Thorne were able
to show that traversable wormholes were compatible with our current
understanding of general relativity and semiclassical quantum
gravity --- but that there was a definite price to be paid --- one
had to admit violations of the null energy condition.

More precisely, what Morris and Thorne showed was equivalent to
the statement that for spherically symmetric traversable wormholes
there must be an open region surrounding the throat over which the
null energy condition is violated~\cite{Morris-Thorne,Visser}.
The striking nature of this result has led numerous authors to try
to find ways of evading or minimizing the energy condition violations,
and on the other hand has led to a number of general theorems
guaranteeing the existence of these violations.

For instance, static but not spherically symmetric thin-shell
wormholes and their variants allow you to move the energy condition
violations around in space, so that there are some routes through the
wormhole that do not encounter energy condition violations. (This is
most easily seen using ``cut and paste'' wormholes constructed using
the thin shell formalism~\cite{Visser89a}. See
also~\cite[pages 153--194]{Visser}.)

Time dependent but spherically symmetric wormholes allow you to
move the energy condition violations around in
time~\cite{Roman,Hochberg-Kephart,Kar,Kar2,Kim}. (Unfortunately
radial null geodesics through the wormhole will still encounter
energy condition violations, subject to suitable technical
qualifications.)

The Friedman--Schleich--Witt topological censorship
theorem~\cite{Topological-censorship}, under suitable technical
conditions, guarantees that spacetimes containing traversable
wormholes must contain some null geodesics that violate the averaged
null energy condition (but gives little information on where these
null geodesics must be).

Attempts at eliminating the energy condition violations completely
typically focus on alternative gravity theories (Brans--Dicke
gravity, Dilaton gravity, gravity with torsion). We argue that such
attempts are at best sleight of hand---it is sometimes possible to
hide the energy condition violations in the Brans--Dicke field, or
the dilaton field, or the torsion, but the energy condition violations
are still always there.

\subsection{Morris--Thorne wormhole}

Take any spherically symmetric static spacetime geometry and use
the radial proper distance as the radial coordinate. The metric
can then without loss of generality be written as

\begin{equation}
ds^2 = - e^{2\phi(l)} dt^2 + dl^2 
       + r^2(l) \left[ d\theta^2 + \sin^2\theta \, d\varphi^2 \right].
\end{equation} 

If we wish this geometry to represent a Morris--Thorne wormhole then
we must impose conditions both on the throat and on asymptotic
infinity~\cite{Morris-Thorne,Visser}.
\begin{itemize}
\item
Conditions at the throat:
\begin{itemize}
\item
The absence of event horizons implies that $\phi(l)$ is everywhere finite.
\item 
The radius of the wormhole throat is defined by
\begin{equation}
r_0 = {\rm min} \{r(l)\}.
\end{equation}
\item
For simplicity one may assume that there is only one such minimum
and that it is an isolated minimum. Generalizing this point is
straightforward.
\item 
Without loss of generality, we can take this throat to occur at
$l=0$.
\item 
The metric components should be at least twice differentiable as
functions of $l$.
\end{itemize}
\item
Conditions at asymptotic infinity:
\begin{itemize}
\item
The coordinate $l$ covers the entire range $(-\infty,+\infty)$.
\item 
There are two asymptotically flat regions, at $l\approx \pm \infty$.
\item 
In order for the spatial geometry to tend to an appropriate
asymptotically flat limit we impose
\begin{equation}
\lim_{l\to\pm\infty}
\{ r(l)/|l| \}= 1.  
\end{equation}
\item 
In order for the spacetime geometry to tend to an appropriate
asymptotically flat limit, we impose finite limits 
\begin{equation}
\lim_{l\to\pm\infty} \phi(l)= \phi_\pm.
\end{equation}
\end{itemize}
\item 
These are merely the minimal requirements to obtain a wormhole that is
``traversable in principle''. For realistic models, ``traversable in
practice'', one should address additional engineering issues such as
tidal effects~\cite[pages 137--152]{Visser}.
\item
We shall argue, later in this survey, that the conditions imposed at
asymptotic infinity can be relaxed (and for certain questions,
asymptotic infinity can be ignored completely), and that it is the
throat of the wormhole that is often more of direct interest.
\end{itemize}
It is an easy exercise to show that the Einstein tensor
is~\cite[pages 149-150]{Visser} 

\begin{eqnarray}
G_{\hat t \hat t} &=& 
-{2r''\over r} + {1-(r')^2\over r^2}, \\
G_{\hat r \hat r} &=& 
{2\phi' r'\over r} - {1-(r')^2\over r^2}, \\
G_{\hat\theta\hat\theta} &=& 
G_{\hat\varphi\hat\varphi} =
\phi'' + (\phi')^2 + {\phi' r' + r'' \over r}.
\end{eqnarray} 
Thus in particular

\begin{equation}
G_{\hat t \hat t} + G_{\hat r \hat r} = -{2r''\over r} + {2\phi' r'\over r}.
\end{equation}
But by definition $r'=0$ at the throat. We also know that $r''\geq
0$ at the throat. The possibility that $r''=0$ at the throat forces
us to invoke some technical complications. By definition the throat
is a local minimum of $r(l)$. Thus there must be an open region
$l\in(0,l^+_*)$ such that $r''(l) > 0$, and on the other side of
the throat an open region $l\in(-l^-_*,0)$ such that $r''(l)>0$.
This is the Morris--Thorne ``flare--out'' condition. 
So for the Einstein tensor we have

\begin{equation}
\exists\; l^-_*, l^+_* > 0 : \qquad
\forall l \in (-l^-_*,0)\cup(0,l^+_*), \qquad
G_{\hat t \hat t} + G_{\hat r \hat r} < 0.
\end{equation}
This constraint on the components of the Einstein tensor follows
directly from the definition of a traversable wormhole and the
definition of the Einstein tensor---it makes no reference to the
dynamics of general relativity and automatically holds {\em by
definition} regardless of whether one is dealing with Einstein
gravity, Brans--Dicke gravity, or any other exotic form of gravity.
As long as you have a spacetime metric that contains a wormhole,
and calculate the Einstein tensor using that {\em same} spacetime
metric, the above inequality holds by definition.

We can always define the total stress energy by enforcing the
Einstein equations

\begin{equation}
G^{\mu\nu} = 8 \pi G \; T_{total}^{\mu\nu}.
\end{equation}{
In terms of the total stress energy 

\begin{equation}
\exists\; l^-_*, l^+_* > 0 : \qquad
\forall l \in (-l^-_*,0)\cup(0,l^+_*), \qquad
T^{total}_{\hat t \hat t} + T^{total}_{\hat r \hat r} < 0.
\end{equation}
Thus the null energy condition for the total stress energy must be
violated on some open region surrounding the throat. The only
requirements for this result are essentially matters of definition:
use the same metric to define the wormhole and to calculate the Einstein
tensor, and use that same Einstein tensor to identify the total
stress-energy.

(This also shows where the maneuvering room is in exotic theories
of gravity. If you have multiple metrics, multiple Einstein tensors,
and multiple definitions of stress-energy then it becomes easy to
hide the violations of the energy conditions in inappropriate
definitions.)

\subsection{Spherically symmetric time-dependent wormholes}

After the initial treatment by Morris and Thorne it was quickly
realized that by eschewing spherical symmetry it is quite possible
to minimize the violations of the null energy
condition~\cite{Visser89a}. In particular it is quite possible
to move the regions subject to energy condition violations around
in space so as to hide them in the woodwork and permit at least
some travelers through the wormhole completely avoid any personal
contact with exotic matter.

Somewhat later, it was realized that {\em time dependence} lets
one move the energy condition violating regions around in
time~\cite{Kar,Kar2}, and so leads to a temporary suspension of
the need for energy condition violations. (For related analyses
see also~\cite{Roman,Hochberg-Kephart,Kim}.) The most direct
presentation of the key results can best be exhibited by taking a
spacetime metric that is conformally related to a zero-tidal force
wormhole by a simple time-dependent but space-independent conformal
factor. Thus

\begin{eqnarray}
ds^2 
&=& 
\Omega(t)^2 \Bigg\{ - dt^2 + {dr^2\over1-b(r)/r} 
+ r^2 \left( d\theta^2 + \sin^2\theta d\phi^2 \right) \Bigg\}.
\end{eqnarray}
(This form of the metric has the advantage that null geodesics
remain null geodesics as $\Omega$ is altered.) It is an easy exercise
to see that

\begin{eqnarray}
G_{\hat t\hat t} 
&=&
+\Omega^{-2} 
\left( {b'(r)\over r^2} + {\dot\Omega^2\over\Omega^2} \right),
\\
G_{\hat\theta \hat\theta} 
&=&
+\Omega^{-2} 
\left( 
{b(r)\over 2r^3} -{b'(r)\over 2r^3} + 
{\dot\Omega^2\over\Omega^2} -2 {\ddot\Omega\over\Omega} 
\right),
\\
G_{\hat r \hat r} 
&=&
-\Omega^{-2} 
\left( 
{b(r)\over r^3} + 2 {\ddot\Omega\over\Omega} - {\dot\Omega^2\over\Omega^2} 
\right).
\end{eqnarray}
In particular, looking along the radial null direction

\begin{equation}
G_{\hat t\hat t} + G_{\hat r \hat r} 
=
\Omega^{-2} 
\left( 
-{b(r)\over r^3} + {b'(r)\over r^3}2  
- 2 {\ddot\Omega\over\Omega} + 4 {\dot\Omega^2\over\Omega^2} 
\right)
\end{equation}
We can rewrite this in terms of the Hubble parameter $H$ and
deceleration parameter $q$, by first relating conformal time $t$ to
comoving time $T$ via $\Omega \, dt = dT$. We then deduce

\begin{equation}
H=\Omega^{-1} \; {d\Omega\over dT} = \Omega^{-2}\;\dot\Omega,
\end{equation}
and 
\begin{equation}
q = 
- \Omega \; {d^2\Omega\over dT^2} \left({d\Omega\over dT}\right)^{-2} 
= 1 - {\Omega\; \ddot\Omega  \over \dot\Omega^2 }.
\end{equation}
Then

\begin{equation}
G_{\hat t\hat t} + G_{\hat r \hat r} 
=
\Omega^{-2} 
\left( 
-{b(r)\over r^3} + {b'(r)\over r^3} \right)
+ 2 H^2 (1+q).  
\end{equation}
So if the universe expands quickly enough we can (temporarily)
suspend the violations of the null energy condition. Note that the
first term above is just the static $\Omega=1$ result rescaled to the
expanded universe.  At the throat of the wormhole this is of order
$-\xi/a^2$ where $a$ is the physical size of the wormhole mouth
and $\xi$ is a dimensionless number typically of order unity unless
some fine tuning is envisaged.  This implies that the expansion of
the universe can overcome the violations of the energy condition
only if

\begin{equation}
a > \sqrt{\xi\over1+q} \; {1\over H}.
\end{equation}
This requires a wormhole throat with radius of order the Hubble
distance or larger (modulo possible fine tuning). This indicates
that suspension of the energy condition violations may be of some
interest during the very early universe when the Hubble parameter is large,
but that in the present epoch it would be quite impractical to rely
on the expansion of the universe to avoid the need for energy
condition violations. 

It is perhaps surprising to realise that the inflationary epoch,
though it can be invoked to make wormholes larger by inflating them
out of the Planck slop (Wheeler's spacetime foam) up to more manageable
sizes~\cite{Roman,Hochberg-Kephart}, cannot be relied upon to aid in
the suspension of energy condition violations. This arises because
$q=-1$ during the inflationary epoch so that for the metric considered
above

\begin{equation}
G_{\hat t\hat t} + G_{\hat r \hat r} 
=
\Omega^{-2} 
\left( 
-{b(r)\over r^3} + {b'(r)\over r^3} \right).  
\end{equation}
This should be compared to equation (3.8) of~\cite{Roman}, which was
obtained directly in terms of a comoving time formalism assuming
inflationary expansion from the outset. In the formalism developed
above, this can be checked by noting that during the inflationary
epoch

\begin{equation}
\Omega = \exp(HT) = {1\over1-Ht},
\end{equation}
where we have normalized $t=0$ at $T=0$. It is then easy to explicitly
check that $q=-1$.

If we want to compare this result with that of the topological
censorship theorem~\cite{Topological-censorship} it is important to
realise that the topological censorship theorem applies to the current
situation only if we switch off the expansion of the universe at
sufficiently early and late times. This is because the topological
censorship theorem uses asymptotic flatness, in both space and time,
as an essential ingredient in setting up both the statement of the
theorem and the proof. (Scri$^-$, past null infinity, simply makes no
sense in a big bang spacetime.)

So, provided we we switch off the expansion of the universe at
sufficiently early and late times, we deduce first that null energy
condition violations reappear at sufficiently early and late times,
and more stringently, that radial null geodesics must violate the
ANEC.

In brief, the suspension of the energy condition violations afforded
by time dependence are either transitory or intimately linked to the
existence of a big bang singularity, and in either case are most
likely limited to small scale microscopic wormholes in the
pre-inflationary epoch.

\subsection{Brans--Dicke wormholes}

Brans--Dicke wormholes~\cite{Agnese,Nandi,Anchordoqui} are particular
examples of the effect that choosing a non-Einstein model for
gravity has on the behaviour of traversable wormholes. The Brans--Dicke
theory of gravity is perhaps the least violent alteration to Einstein
gravity that can be contemplated. In the Brans--Dicke theory the
gravitational field is composed of two components: a spacetime
metric plus a dynamical scalar field.  (A word of caution: some
papers dealing with Brans--Dicke wormholes are actually in Euclidean
signature~\cite{Yang-Zhang}. Euclidean wormholes are
qualitatively different from Lorentzian wormholes and will not be
discussed in this survey.)

\subsubsection{The Jordan Frame}

In the so-called Jordan frame the Action is~\cite[page 1070]{MTW}

\begin{equation}
S = \int\sqrt{-g} 
\left\{ 
\phi R - \omega {(\nabla\phi)^2\over\phi} + 16\pi G \;{\cal L}_{matter} 
\right\}.
\end{equation}
The equations of motion are

\begin{equation}
G_{\alpha\beta} = 
{8\pi G \over\phi} T^{matter}_{\alpha\beta} 
+{\omega\over\phi^2} 
\left\{ 
\phi_\alpha \phi_\beta - {1\over2} g_{\alpha\beta} (\nabla\phi)^2 
\right\}
+ {1\over\phi} 
\left\{ 
\phi_{;\alpha\beta} - g_{\alpha\beta} \nabla^2  \phi 
\right\},
\end{equation}
and

\begin{equation}
\nabla^2 \phi = {8\pi G\over3+2\omega} \; T.
\end{equation}
If we wish the metric $g$ to describe a spherically symmetric static
wormhole, then the Morris-Thorne analysis implies that the total
stress energy defined by $T^{total}_{\alpha\beta} =
G_{\alpha\beta}/(8\pi G)$ must violate the null energy
condition. Whether or not the ``matter'' part of the total
stress-energy violates the null energy condition depends on how the
Brans--Dicke scalar field behaves at and near the throat.

If we assume that the wormholes we are looking for have $\phi\neq0$
and $\phi\neq\infty$ at the throat, then for any radial null vector
$k^\alpha$ we have

\begin{equation}
G_{\alpha\beta} \; k^\alpha \; k^\beta = 
{8\pi G \over\phi} \; T^{matter}_{\alpha\beta}  \; k^\alpha \; k^\beta
+ \omega { (k^\alpha \phi_\alpha)^2 \over \phi^2} 
+ {\phi_{;\alpha\beta} \; k^\alpha \; k^\beta \over\phi} \leq 0. 
\end{equation}
This implies that the only way in which the ``matter'' part of the of
the stress-energy can avoid violating the null energy condition is if
either $\omega<0$ and $\nabla\phi\neq0$ at the throat, or if $\phi$ is
convex at the throat. We now verify these general conclusions by
looking at some specific exact solutions.

\subsubsection{Vacuum Brans--Dicke wormholes [Jordan frame]}

For vacuum Brans--Dicke gravity a suitably large class of solutions to
the field equations is~\cite{Agnese}

\begin{eqnarray}
ds^2 
&=&
-\left[ {1-R/r \over 1+R/r} \right]^{2A} dt^2
\nonumber
\\
&&
+\left[ 1+R/r \right]^4
\left[ 
{1-R/r \over 1+R/r } 
\right]^{2+2B}
\left[ dr^2 + r^2 \left(d\theta^2 + \sin^2\theta d\phi^2 \right) \right].
\end{eqnarray}

\begin{equation}
\phi = \phi_0 
\left[ 
{1-B/r \over 1+B/r } 
\right]^{-(A+B)}.
\end{equation}
(Note there is a non-propagating typo in equation (8) of~\cite{Agnese}.)
Here we have chosen to work in isotropic coordinates. We have

\begin{equation}
R = \sqrt{3+2\omega\over4+2\omega} \; {G\;M\over2}.
\end{equation}
(This is necessary to get the correct asymptotic behaviour as $r \to
\infty$.) We also have the constraints that

\begin{equation}
\phi_0 = {4+2\omega\over3+2\omega}.
\end{equation}

\begin{equation}
A = \sqrt{4+2\omega\over3+2\omega} > 0.
\end{equation}

\begin{equation}
B = - {1+\omega\over2+\omega} \sqrt{4+2\omega\over3+2\omega}.
\end{equation}

The metric is real only for $\omega>-3/2$ or $\omega<-2$, the square
root always being taken to be positive when it is real, and the metric
reduces to the Schwarzschild geometry for $\omega\to\pm\infty$. The
geometry also has a symmetry under $r \to R^2/r$, but this is
potentially misleading. The surface $r=R$ is {\em not} the throat of a
wormhole. To see what is going on, consider the proper circumference
of a circle at radius $r$ circumscribing the geometry. we have

\begin{equation}
C(r) = 2 \pi r 
\left[ 1+R/r \right]^2
\left[ 
{1-R/r \over 1+R/r } 
\right]^{1+B}.
\end{equation}

Assume that $R$ (and hence $M$) is positive, this can be generalized
if desired.

\begin{itemize}
\item
If $B>-1$ then $C(r) \to 0$ as $r \to R$. In this case the
surface $r=R$ is a naked curvature singularity as may be verified by
direct computation of the curvature invariants. The region
$r\in(0,R)$ is a second asymptotically flat region, isomorphic to the
first, that connects to the first only at the curvature singularity
$r=R$. [$B>-1$  corresponds to $\omega\in(-3/2,+\infty)$.]
\item
If $B=-1$ then $C(r) \to 8\pi R$ as $r \to R$. In this case the
surface $r=R$ is at least a surface of finite area. In fact $B=-1$ is
achieved only for $\omega=\pm\infty$ in which case the geometry
reduces to Schwarzschild. (In this case we also have $A=1$ and
$\phi=\phi_0$.)
\item
If $B<-1$ then $C(r) \to \infty$ as $r \to R$. This corresponds to
$\omega\in(-\infty,-2)$. In this case the surface $r=R$ is the second
asymptotic spatial infinity associated with a traversable
wormhole. The location of the wormhole throat is specified by looking
for the minimum value of $C(r)$ which occurs at
\begin{equation}
r_{throat} = 
R \left[ 
-B+\sqrt{B^2-1}
\right]>R.
\end{equation}
The wormhole is in this case {\em asymmetric} under interchange of the
two asymptotic regions ($r=\infty$ and $r=R$). Since $A+B<0$ in this
parameter regime, $\phi\to0$ as $r\to R$, and so the effective Newton
constant $G_{eff} = G/\phi$ tends to infinity on the other side of the
wormhole. The region near $r=R$ is asymptotically large, but not
asymptotically flat, as may be verified by direct computation of the
curvature. (It would be interesting to know a little bit more about
what this region actually looks like, and to develop a better
understanding of the physics on the other side of this class of
Brans--Dicke wormholes.)  The region $r\in(0,R)$ is now a second
completely independent universe, isomorphic to the first, with its own
wormhole occurring at
\begin{equation}
r_{throat} = 
R \left[ 
-B-\sqrt{B^2-1}
\right]<R.
\end{equation}
\item
There is even more parameter space to explore if we look at the
extended class of Brans--Dicke solutions discussed in~\cite{Nandi}, or
let the total mass go negative.
\end{itemize}

In summary, for suitable choices of the parameters, $\omega < -2$,
there are wormholes in vacuum Brans--Dicke gravity expressed in the
Jordan frame. The null energy condition is still violated, with the
Brans--Dicke field providing the exotic matter.

\subsubsection{The Einstein Frame}

By making a conformal transformation it is possible to express the
Brans--Dicke theory in the so-called Einstein frame. The action is now
(see~\cite{Yang-Zhang}, modified for Lorentzian signature)

\begin{equation}
S = \int\sqrt{-\tilde g} 
\left\{ 
{R(\tilde g)\over2\omega+3} +
{1\over2} (\nabla\sigma)^2+ 16\pi G \;\exp(2\sigma){\cal L}_{matter} 
\right\}.
\end{equation}

Here

\begin{equation}
\tilde g_{\mu\nu} = \exp(\sigma) \; g_{\mu\nu}.
\end{equation}

\begin{equation}
\phi = {1\over(2\omega+3)} \exp(\sigma).
\end{equation}
(And note that the matter Lagrangian, which implicitly depends on the
Jordan metric, also has to be carefully rewritten in terms of $\sigma$
and the Einstein metric.) Now provided $\phi$ is neither zero nor
infinite the conformal transformation from the Jordan to the Einstein
frame is globally well-defined. If in addition $\phi$ goes to a finite
non-zero constant at in the asymptotically flat region then it is
clear that wormholes, in the sense of topologically nontrivial curves
from Scri$^-$ to Scri$^+$ (past null infinity to future null infinity)
can neither be created no destroyed by a change of frame.  (For
background see~\cite{Topological-censorship} and~\cite[pages
195--199]{Visser}).  What can, and in general does, change however is
the location and even the number of wormhole throats that are
encountered in crossing from one asymptotically flat region to
another.  A wormhole throat as defined by the Einstein frame metric is
not necessarily a wormhole throat as defined by the Jordan frame
metric.

\subsubsection{Vacuum Brans--Dicke wormholes [Einstein frame]}

If we look at the vacuum Brans--Dicke field equations in the Einstein
frame we get
\begin{equation}
G_{\alpha\beta} =  (2\omega+3)
\left\{ 
\sigma_\alpha \sigma_\beta - {1\over2} g_{\alpha\beta} (\nabla\sigma)^2 
\right\},
\end{equation}
and

\begin{equation}
\nabla^2 \sigma = 0.
\end{equation}
In the Einstein frame, the Brans--Dicke field violates the null energy
condition only for $\omega<-3/2$, and so vacuum Brans--Dicke wormholes
can exist only for $\omega<-3/2$. (This is a necessary but not
sufficient condition.) Note that the Jordan and Einstein frames are
globally conformally equivalent only if $\phi$ is never zero. Since
the $\omega<-2$ Jordan frame wormholes have $\phi=0$ at $r=R$ there is
a risk of pathological behaviour.

If we take the class of vacuum Brans--Dicke solutions considered
previously, and transform them to the Einstein frame we have

\begin{eqnarray}
(2\omega+4) d\tilde s^2 
&=&
-\left[ {1-R/r \over 1+R/r} \right]^{A-B} dt^2
\nonumber
\\
&&
+\left[ 1+R/r \right]^4
\left[ 
{1-R/r \over 1+R/r } 
\right]^{2+B-A}
\left[ dr^2 + r^2 \left(d\theta^2 + \sin^2\theta d\phi^2 \right) \right].
\nonumber
\\
&&
\end{eqnarray}
The proper circumference of a circle at radius $r$ circumscribing the
geometry is now

\begin{equation}
\tilde C(r) = {2 \pi r \over\sqrt{|2\omega+4|}} 
\left[ 1+R/r \right]^2
\left[ 
{1-R/r \over 1+R/r } 
\right]^{1+(B-A)/2}.
\end{equation}
Assume again that $R$ (and hence $M$) is positive. It is now the
combination $B-A$ that is of central interest. Indeed

\begin{equation}
A-B =  2 \sqrt{3+2\omega\over4+2\omega} >0.
\end{equation}

\begin{itemize}
\item
If $B-A>-2$ then $\tilde C(r) \to 0$ as $r \to R$. This again
corresponds to $\omega\in(-3/2,+\infty)$, and the geometry again
contains a naked curvature singularity as may be verified by direct
computation of the curvature invariants. The region $r\in(0,B)$ is a
second asymptotically flat region, isomorphic to the first, that
connects to the first only at the curvature singularity $r=B$.
\item
If $B-A=2$ then $\tilde C(r) \to 8\pi R$ as $r \to R$. In this case
the surface $r=R$ is at least a surface of finite area. It is easy to
see that $B-A=-2$ implies $\omega=\pm\infty$ and thus $A=1$, $B=-1$.
This again reproduces the Schwarzschild solution.
\item
If $B-A<-2$ then $\tilde C(r) \to
\infty$ as $r \to R$. The surface $r=R$ is the second
asymptotic spatial infinity associated with a traversable
wormhole. The location of the wormhole throat is specified by looking
for the minimum value of $\tilde C(r)$ which occurs at
\begin{equation}
r_{throat} = 
R \left[ 
{(A-B)\over2} + \sqrt{{(A-B)^2\over4} - 1}
\right].
\end{equation}
The wormhole is again {\em asymmetric} under interchange of the two
asymptotic regions ($r=\infty$ and $r=R$), and the throat is located
at a different place.
\item
In some sense we have been lucky: The field equations have forced the
conformal transform that relates the Jordan and Einstein frames to be
sufficiently mild that the two frames agree as to the range of values
of the $\omega$ parameter that lead to traversable wormhole
geometries. There is no a priori necessity for this agreement. (In
fact when considering $O(4)$ Euclidean Brans--Dicke wormholes these
differences are rather severe~\cite{Yang-Zhang}.) The two frames do
differ however in the precise location of the wormhole throat, and in
the precise details of the energy condition violations. In the
Einstein frame the energy condition violations are obvious from the
relative minus sign (for $\omega < -3/2$) in front of the kinetic
energy term for the $\sigma$ field. In the Jordan frame the energy
condition violations are still encoded in the Brans--Dicke field (now
$\phi$) but in a more subtle manner.
\end{itemize}

In summary, for suitable choices of the parameters ($\omega<-2$), there
are wormholes in vacuum Brans--Dicke gravity expressed in the Einstein
frame. The null energy condition is still violated, with the
Brans--Dicke field providing the exotic matter.

\subsubsection{Other exotic wormholes}

Similar analyses can be performed for other more or less natural
generalizations of Einstein gravity. Dilaton gravity is a particularly
important example, inspired by string theory, that is very closely
related to Brans-Dicke gravity. Other examples include Einstein--Cartan
gravity, Lovelock gravity, Gauss--Bonnet gravity, higher-derivative
gravity, etc...

\section{Generic static wormholes}

We now set aside the special cases we have been discussing, and seek
to develop a general analysis of the energy condition violations in
static wormholes. (This analysis is based largely
on~\cite{Hochberg-Visser}. For an analysis using similar techniques
applied to static vacuum and electrovac black holes see
Israel~\cite{Israel67,Israel68}. A related decomposition applied to
the collapse problem is addressed in~\cite{Israel86}.) In view of the
preceding discussion we want to get away from the notion that
topology is the intrinsic defining feature of wormholes and instead
focus on the geometry of the wormhole throat. Our strategy is
straightforward
\begin{itemize}
\item
Take any static spacetime, and use the natural time coordinate to
slice it into space plus time.
\item
Use the Gauss--Codazzi and Gauss--Weingarten equations to
decompose the $(3+1)$--dimensional spacetime curvature tensor in
terms of the $3$--dimensional spatial curvature tensor, the extrinsic
curvature of the time slices [zero!], and the gravitational potential.
\item
Take any $3$--dimensional spatial slice, and look for a $2$-dimensional
surface of strictly minimal area. Define such a surface, if it
exists, to be the throat of a wormhole. This generalizes the
Morris--Thorne flare out condition to arbitrary static wormholes.
\item
Use the Gauss--Codazzi and Gauss--Weingarten equations again, this
time to decompose the $3$--dimensional spatial curvature tensor in
terms of the $2$--dimensional curvature tensor of the throat and
the extrinsic curvature of the throat as an embedded hypersurface
in the $3$--geometry.
\item
Reassemble the pieces: Write the spacetime curvature in terms of
the $2$ curvature of the throat, the extrinsic curvature of the
throat in $3$-space, and the gravitational potential.
\item
Use the generalized flare-out condition to place constraints on
the stress-energy at and near the throat.
\end{itemize}

\subsection{Static spacetimes}

In any static spacetime one can decompose the spacetime metric
into block diagonal form~\cite{MTW,Hawking-Ellis,Wald}:

\begin{eqnarray}
ds^2 &=& g_{\mu\nu} \; dx^\mu dx^\nu
\\
&=&
- \exp(2\phi) dt^2 + g_{ij} \; dx^i dx^j.
\end{eqnarray}

\noindent
{\em Notation:} Greek indices run from 0--3 and refer to space-time;
latin indices from the middle of the alphabet ($i$, $j$, $k$, \dots) run
from 1--3 and refer to space; latin indices from the beginning of
the alphabet ($a$, $b$, $c$, \dots) will run from 1--2 and will be used
to refer to the wormhole throat and directions parallel to the
wormhole throat.

Being static tightly constrains the space-time geometry in terms
of the three-geometry of space on a constant time slice, and the
manner in which this three-geometry is embedded into the spacetime.
For example, from \cite[page 518]{MTW} we have

\begin{eqnarray}
{}^{(3+1)}R_{ijkl} &=& 
{}^{(3)}R_{ijkl}. 
\\
{}^{(3+1)}R_{\hat tabc} &=& 0.
\\
{}^{(3+1)}R_{\hat ti\hat tj}  &=&  
\phi_{|ij} + \phi_{|i} \; \phi_{|j}.
\end{eqnarray}

\noindent
The hat on the $t$ index indicates that we are looking at components
in the normalized $t$ direction

\begin{equation}
X_{\hat t} = X_t \sqrt{-g^{tt}} = X_t \; \exp(-\phi).
\end{equation}

\noindent
This means we are using an orthonormal basis attached to the fiducial
observers (FIDOS). We use $X_{;\alpha}$ to denote a space-time
covariant derivative; $X_{|i}$ to denote a three-space covariant
derivative, and will shortly use $X_{:a}$ to denote two-space
covariant derivatives taken on the wormhole throat itself.

Now taking suitable contractions,

\begin{eqnarray}
{}^{(3+1)}R_{ij} &=& 
{}^{(3)}R_{ij} -  \phi_{|ij} - \phi_{|i} \; \phi_{|j}.
\\
{}^{(3+1)}R_{\hat t i} &=&  0 .
\\
{}^{(3+1)}R_{\hat t\hat t} &=&  g^{ij}
\left[  
\phi_{|ij} + \phi_{|i} \phi_{|j}
\right].
\end{eqnarray}

\noindent
So

\begin{equation}
{}^{(3+1)}R =  {}^{(3)}R - 2 g^{ij}
\left[ \phi_{|ij} + \phi_{|i} \phi_{|j} \right].
\end{equation}

\noindent
To effect these contractions, we make use of the decomposition
of the spacetime metric in terms of the spatial three-metric, the
set of vectors ${e^{\mu}_i}$ tangent to the time-slice, and the
vector $V^\mu = \exp[\phi] \; ({\partial}/{\partial t } )^{\mu}$
normal to the time slice: 

\begin{equation}
{}^{(3+1)}g^{\mu \nu} = e^{\mu}_i e^{\nu}_j \; g^{ij} - V^\mu V^\nu.
\end{equation}
Finally, for the spacetime Einstein tensor (see~\cite[page
552]{MTW})

\begin{eqnarray}
\label{E-static-stress-energy-b}
{}^{(3+1)}G_{ij} &=&  {}^{(3)}G_{ij}
 -  \phi_{|ij} - \phi_{|i} \; \phi_{|j}
+ g_{ij} \; g^{kl}\left[ \phi_{|kl} +\phi_{|k} \phi_{|l} \right].
\\
{}^{(3+1)}G_{\hat t i} &=& 0 .
\\
{}^{(3+1)} G_{\hat t\hat t} &=& + {1\over2} {}^{(3)} R.
\label{E-static-stress-energy-e}
\end{eqnarray}

\noindent
This decomposition is generic to {\em any} static spacetime.  (You
can check this decomposition against various standard textbooks to
make sure the coefficients are correct. For instance see Synge~\cite[page
339]{Synge}, Fock~\cite{Fock}, or
Adler--Bazin--Schiffer~\cite{Adler-Bazin-Schiffer})

{\em Observation:} Suppose the strong energy condition (SEC) holds
then~\cite{Visser}

\begin{eqnarray}
\label{E-static-SEC-b}
SEC &\implies& (\rho + g_{ij} T^{ij}) \geq 0
\\
&\implies& g^{ij}
\left[  
\phi_{|ij} + \phi_{|i} \phi_{|j}
\right] \geq 0
\\
&\implies& \hbox{$\phi$ has no isolated maxima}.
\label{E-static-SEC-e}
\end{eqnarray}
This is a nice consistency check, and also helpful in understanding
the physical import of the strong energy condition.

\subsection{Definition of a generic static throat}

We define a traversable wormhole throat, $\Sigma$, to be a
$2$--dimensional hypersurface of {\em minimal} area taken in one of
the constant-time spatial slices. Compute the area by taking

\begin{equation}
A(\Sigma) = \int \sqrt{{}^{(2)}g} \; d^{2} x.
\end{equation}

\noindent
Now use Gaussian normal coordinates, $x^i=(x^a;n)$, wherein the
hypersurface $\Sigma$ is taken to lie at $n=0$, so that

\begin{equation}
{}^{(3)}g_{ij} \; dx^i dx^j  = {}^{(2)}g_{ab} \; dx^a dx^b  +  dn^2.
\end{equation}

\noindent
The variation in surface area, obtained by pushing the surface $n=0$
out to $n = \delta n(x)$, is given by the standard computation

\begin{equation}
\delta A(\Sigma) = 
\int {\partial\sqrt{{}^{(2)}g}\over \partial n} \; 
\delta n(x) \; d^{2} x.
\end{equation}

\noindent
Which implies

\begin{equation}
\delta A(\Sigma) =
\int \sqrt{{}^{(2)}g} \;
{1\over2} \; g^{ab} \; {\partial g_{ab} \over \partial n} \; 
\delta n(x) \; d^{2} x.
\end{equation}

\noindent
In Gaussian normal coordinates the extrinsic curvature is simply
defined by

\begin{equation}
K_{ab} = - {1\over2} {\partial g_{ab} \over \partial n}.
\end{equation}

\noindent
(See \cite[page 552]{MTW}. We use MTW sign conventions. The convention
in \cite[page 156]{Visser} is opposite.) Thus

\begin{equation}
\delta A(\Sigma) =  - \int \sqrt{{}^{(2)}g} \; \tr(K) \; 
\delta n(x) \; d^{2} x.
\end{equation}

\noindent
[We use the notation $\tr(X)$ to denote $g^{ab} \; X_{ab}$.] Since
this is to vanish for arbitrary $\delta n(x)$, the condition that
the area be {\em extremal} is simply $\tr(K)=0$.  To force the area
to be {\em minimal} requires (at the very least) the additional
constraint $\delta^2 A(\Sigma) \geq 0$. (We shall also consider
higher-order constraints below.)  But by explicit calculation

\begin{equation}
\delta^2 A(\Sigma) =  
- \int \sqrt{{}^{(2)}g} \;
\left( {\partial\tr(K)\over \partial n} - \tr(K)^2 \right) \; 
\delta n(x) \; \delta n(x) \; d^{2} x.
\end{equation}

\noindent
Extremality [$\tr(K)=0$] reduces this minimality constraint to

\begin{equation}
\delta^2 A(\Sigma) =  
- \int \sqrt{{}^{(2)}g} \;
\left( {\partial\tr(K)\over \partial n} \right) \; 
\delta n(x) \; \delta n(x) \; d^{2} x \geq 0.
\end{equation}

\noindent
Since this is to hold for arbitrary $\delta n(x)$ this implies that
at the throat we certainly require

\begin{equation}
{\partial\tr(K)\over \partial n} \leq 0.
\end{equation}

\noindent
This is the generalization of the Morris--Thorne ``flare-out''
condition to arbitrary static wormhole throats.

We now invoke some technical fiddles related to the fact that we
eventually prefer to have a strong inequality $(<)$ at or near the
throat, in preference to a weak inequality $(\leq)$ at the throat.
Exactly the same type of technical fiddle is required when considering
the Morris--Thorne spherically symmetric wormhole.  In the following
definitions, the two-surface referred to is understood to be embedded
in a three-dimensional space, so that the concept of its extrinsic
curvature (relative to that embedding space) makes sense.

\noindent
{\bf Definition: Simple flare-out condition.}\\ 
{\em A two-surface satisfies the ``simple flare-out'' condition if and
only if it is extremal, $\tr(K)=0$, and also satisfies
${\partial\tr(K)/\partial n} \leq 0$.}

This flare-out condition can be rephrased as follows: We have as
an identity that

\begin{equation}
{\partial\tr(K)\over \partial n} = 
\tr\left({\partial K\over \partial n}\right) + 2 \tr(K^2).
\end{equation}

\noindent
So minimality implies

\begin{equation}
\tr\left({\partial K\over \partial n}\right) + 2 \tr(K^2) \leq 0.
\end{equation}

\begin{figure}[htb]
\vbox{\hfil\epsfbox{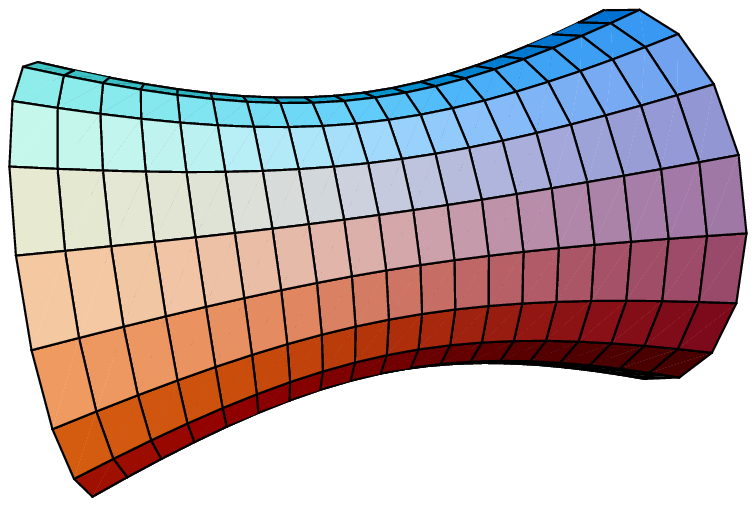}\hfil}
\caption[Generic wormhole throat]
{\label{F-generic}
Generically, we define the throat to be located at a true minimum
of the area. The geometry should flare-out on either side of the
throat, but we make no commitment to the existence of any asymptotically
flat region.}
\end{figure}

\noindent
Generically we would expect the inequality to be strict, in the
sense that ${\partial\tr(K) / \partial n} < 0$, for at least some
points on the throat. (See figure~\ref{F-generic}.) This suggests
the modified definition below.

\noindent
{\bf Definition: Strong flare-out condition.}\\
{\em A two surface satisfies the ``strong flare-out'' condition
at the point $x$ if and only if it is extremal, $\tr(K)=0$,
everywhere satisfies ${\partial\tr(K)/\partial n} \leq 0$, and 
if at the point $x$ on the surface the inequality is strict:}

\begin{equation}
{\partial\tr(K)\over\partial n} < 0.
\end{equation}

It is sometimes sufficient to demand a weak integrated form of the
flare-out condition. 

\noindent
{\bf Definition: Weak flare-out condition.}\\ 
{\em A two surface satisfies the ``weak flare-out'' condition if
and only if it is extremal, $\tr(K)=0$, and}

\begin{equation}
\int \sqrt{{}^{(2)}g} \;{\partial\tr(K)\over \partial n} d^2 x < 0.
\end{equation}

Note that the strong flare-out condition implies both the simple
flare-out condition and the weak flare-out condition, but that the
simple flare-out condition does not necessarily imply the weak
flare-out condition. (The integral could be zero.) Whenever
we do not specifically specify the type of flare-out condition
being used we deem it to be the simple flare-out condition.

The conditions under which the weak definition of flare-out are
appropriate arise, for instance, when one takes a Morris--Thorne
traversable wormhole (which is symmetric under interchange of the
two universes it connects) and distorts the geometry by placing a
small bump on the original throat. (See figure~\ref{F-bump}.)

\begin{figure}[htb]
\vbox{\hfil\epsfbox{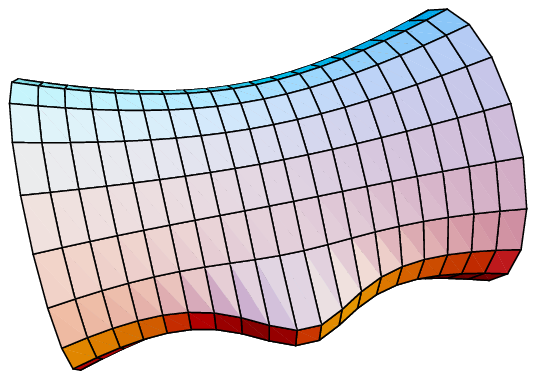}\hfil}
\caption[Strong versus weak throats]
{\label{F-bump}
Strong versus weak throats: Placing a small bump on a strong throat
typically causes it to tri-furcate into two strong throats plus a
weak throat.}
\end{figure}

The presence of the bump causes the old throat to trifurcate into
three extremal surfaces: Two minimal surfaces are formed, one on
each side of the old throat, (these are minimal in the strong sense
previously discussed), while the surface of symmetry between the
two universes, though by construction still extremal, is no longer
minimal in the strict sense. However, the surface of symmetry is
often (but not always) minimal in the weak (integrated) sense
indicated above. 

A second situation in which the distinction between strong and weak
throats is important is in the cut-and-paste construction for
traversable wormholes~\cite{Visser,Visser89a,Visser89b}. In this
construction one takes two (static) spacetimes (${\cal M}_1$, ${\cal
M}_2$) and excises two geometrically identical regions of the form
$\Omega_i\times\Re$, $\Omega_i$ being compact spacelike surfaces
with boundary and $\Re$ indicating the time direction. One then
identifies the two boundaries $\partial\Omega_i \times \Re$ thereby
obtaining a single manifold (${\cal M}_1 \# {\cal M}_2$) that
contains a wormhole joining the two regions ${\cal M}_i - \Omega_i
\times \Re$.  We would like to interpret the junction
${\partial\Omega_{1=2} \times \Re}$ as the throat of the wormhole.

If the sets $\Omega_i$ are convex, then there is absolutely no
problem: the junction ${\partial\Omega_{1=2} \times \Re}$ is by
construction a wormhole throat in the strong sense enunciated above.

On the other hand, if the $\Omega_i$ are concave, then it is
straightforward to convince oneself that the junction $\partial\Omega_{1=2}
\times \Re$ is not a wormhole throat in the strong sense.  If one
denotes the {\em convex hull} of $\Omega_i$ by $\conv(\Omega_i)$
then the {\em two} regions $\partial[\conv(\Omega_i)] \times \Re$
{\em are} wormhole throats in the strong sense. The junction
${\partial\Omega_{1=2} \times \Re}$ is at best a wormhole throat
in the weak sense.

For these reasons it is useful to have this notion of a weak throat
available as an alternative definition.  Whenever we do not qualify
the notion of wormhole throat it will refer to a throat in the simple
sense. Whenever we refer to a throat in the weak sense or strong
senses we will explicitly say so. Finally, it is also useful to define

\noindent
{\bf Definition: Weak $f$-weighted flare-out condition}\\ 
{\em A two surface satisfies the ``weak $f$-weighted flare-out'' condition
if and only if it is extremal, $\tr(K)=0$, and}

\begin{equation}
\int \sqrt{{}^{(2)}g} \; f(x) \; 
{\partial\tr(K)\over \partial n} d^2 x < 0.
\end{equation}

\noindent
(We will only be interested in this condition for $f(x)$ some positive
function defined over the wormhole throat.)

The constraints on the extrinsic curvature embodied in these various
definitions lead to constraints on the spacetime geometry, and
consequently constraints on the stress-energy.

\noindent
{\bf Technical point I: Degenerate throats}\\
A class of wormholes for which we have to extend these
definitions arises when the wormhole throat possesses an accidental
degeneracy in the extrinsic curvature at the throat. The previous
discussion has tacitly been assuming that near the throat we can
write

\begin{eqnarray}
{}^{(2)} g_{ab}(x,n) &=& 
{}^{(2)} g_{ab}(x,0) + 
n \left.{\partial\left[{}^{(2)} g_{ab}(x,n)\right]
\over \partial n}\right|_{n=0} 
\nonumber\\
&& \qquad
+ {n^2\over2} 
\left.{\partial^2\left[{}^{(2)} g_{ab}(x,n)\right]
\over (\partial n)^2}\right|_{n=0} +
O[n^3].
\end{eqnarray}

\noindent
with the linear term having trace zero (to satisfy extremality) and the
quadratic term being constrained by the flare-out conditions. 

Now if we have an accidental degeneracy with the quadratic (and
possibly even higher order terms) vanishing identically, we would
have to develop an expansion such as

\begin{eqnarray}
{}^{(2)} g_{ab}(x,n) &=& 
{}^{(2)} g_{ab}(x,0) + 
n \left.{\partial\left[{}^{(2)} g_{ab}(x,n)\right]
\over \partial n}\right|_{n=0} 
\nonumber\\
&& \qquad 
+ {n^{2N}\over{(2N)!}} \;
\left.{\partial^{2N}\left[{}^{(2)} g_{ab}(x,n)\right]
\over (\partial n)^{2N}}\right|_{n=0} +
O[n^{2N+1}].
\end{eqnarray}

\noindent
Applied to the metric determinant this implies an expansion such as

\begin{eqnarray}
\label{E-k-n}
\sqrt{{}^{(2)} g(x,n)} &=& 
\sqrt{{}^{(2)} g(x,0)} 
\left( 
1 + {n^{2N} \; k_N(x) \over {(2N)!}}  + O[n^{2N+1}] 
\right).
\end{eqnarray}

\noindent
Where $k_N(x)$ denotes the first non-zero sub-dominant term 
in the above expansion, and we know by explicit construction that
\begin{eqnarray}
k_N(x) 
&=&
+{1\over2}
\tr \left(\left.
{\partial^{2N}\left[{}^{(2)} g_{ab}(x,n)\right] \over (\partial n)^{2N}}
\right|_{n=0}\right)
\\
&=& 
-\tr \left(\left.  
\frac{\partial^{2N-1} K_{ab}(x,n)}{(\partial n)^{2N-1}}
\right|_{n=0}\right) 
\\
&=& 
-\left(\left. 
\frac{\partial^{2N-1} K(x,n)}{(\partial n)^{2N-1}}
\right|_{n=0}\right),
\end{eqnarray}
since the trace is taken with ${}^{(2)} g^{ab}(x,0) $ and this
commutes with the normal derivative.  We know that the first non-zero
subdominant term in the expansion (\ref{E-k-n}) must be of even
order in $n$ ({\em i.e.} $n^{2N}$), and cannot correspond to an
odd power of $n$, since otherwise the throat would be a point of
inflection of the area, not a minimum of the area. Furthermore,
since $k_N(x)$ is by definition non-zero the flare-out condition
must be phrased as the constraint $k_N(x) > 0$, with this now being
a strict inequality. More formally, this leads to the definition
below.

\noindent
{\bf Definition: $N$-fold degenerate flare-out condition:}\\ {\em
A two surface satisfies the ``$N$-fold degenerate flare-out''
condition at a point $x$ if and only if it is extremal, $\tr(K)=0$,
if in addition the first $2N-2$ normal derivatives of the trace of
the extrinsic curvature vanish at $x$, and if finally at the point
$x$ one has}

\begin{equation}
{\partial^{2N-1}\tr(K)\over (\partial n)^{2N-1} } < 0,
\end{equation}

\noindent
{\em where the inequality is strict. (In the previous notation this
is equivalent to the statement that $k_N(x) > 0$.) }

Physically, at an $N$-fold degenerate point, the wormhole
throat is seen to be extremal up to order $2N-1$ with respect to
normal derivatives of the metric, i.e., the flare-out property is
delayed spatially with respect to throats in which the flare-out
occurs at second order in $n$. The way we have set things up,  the
$1$-fold degenerate flare-out condition is completely equivalent
to the strong flare-out condition.

If we now consider the extrinsic curvature directly we see, by
differentiating (\ref{E-k-n}), first that

\begin{eqnarray}
K(x,n) &=&  - {n^{2N-1} \; k_N(x) \over(2N-1)!}  + O[n^{2N}],
\end{eqnarray}

\noindent
and secondly that

\begin{eqnarray}
{\partial K(x,n) \over\partial n} &=& 
- {n^{2N-2} \; k_N(x) \over(2N-2)!} + O[n^{2N-1}].
\end{eqnarray}

\noindent 
{From} the dominant $n\to 0$ behaviour we see that if (at some
point $x$) $2N$ happens to equal $2$, then the flare-out condition
implies that ${\partial K(x,n) /\partial n}$ must be negative {\em
at and near the throat}. This can also be deduced directly from
the equivalent strong flare-out condition: if  ${\partial K(x,n)
/\partial n}$ is negative and non-zero at the throat, then it must
remain negative in some region surrounding the throat.  On the
other hand, if $2N$ is greater than $2$ the flare-out condition
only tells us that ${\partial K(x,n) /\partial n}$ must be negative
{\em in some region surrounding the the throat}, and does not
necessarily imply that it is negative at the throat itself. (It
could merely be zero at the throat.)

Thus for degenerate throats, the flare-out conditions should
be rephrased in terms of the first non-zero normal derivative beyond
the linear term. Analogous issues arise even for Morris--Thorne
wormholes~\cite[page 405, equation (56)]{Morris-Thorne}, see also
the discussion presented in \cite[pages  104--105, 109]{Visser}.
Even if the throat is non-degenerate ($1$-fold degenerate) there
are technical advantages to phrasing the flare-out conditions this
way:  It allows us to put constraints on the extrinsic curvature
near but not on the throat.

\noindent
{\bf Technical point II: Hyperspatial tubes}\\
A second class of wormholes requiring even more technical
fiddles arises when there is a central section which is completely
uniform and independent of $n$. [So that $K_{ab}=0$ over the whole
throat for some finite range $n\in(-n_0,+n_0)$.] This central
section might be called a ``hyperspatial tube''. The flare-out
condition should then be rephrased as stating that whenever extrinsic
curvature first deviates from zero [at some point $(x,\pm n_0)$]
one must formulate constraints such as

\begin{equation}
\left.{\partial\tr(K) \over \partial n}\right|_{\pm n_0^\pm} \leq 0. 
\end{equation}
In this case $\tr(K)$ is by definition not an analytic function of
$n$ at $n_0$, so the flare-out constraints have to be interpreted
in terms of one-sided derivatives in the region outside the
hyperspatial tube.  [That is, we are concerned with the possibility
that $\sqrt{g(x,n)}$ could be constant for $n< n_0$ but behave as
$(n-n_0)^{2N}$ for $n>n_0$. In this case derivatives, at $n=n_0$,
do not exist beyond order $2N$.]

\subsection{Geometry of a generic static throat}

Using Gaussian normal coordinates in the region surrounding the throat

\begin{equation}
{}^{(3)}R_{abcd} = {}^{(2)}R_{abcd} - (K_{ac} K_{bd} - K_{ad} K_{bc} ).
\end{equation}

\noindent 
See \cite[page 514, equation (21.75)]{MTW}.
Because two dimensions is special this reduces to:

\begin{equation}
{}^{(3)}R_{abcd} = 
{{}^{(2)}R\over 2} \; (g_{ac} g_{bd} - g_{ad} g_{bc} ) 
                    - (K_{ac} K_{bd} - K_{ad} K_{bc} ).
\end{equation}

\noindent
Of course we still have the standard dimension-independent results
that:

\begin{equation}
{}^{(3)}R_{nabc} =  - (K_{ab:c}  - K_{ac:b}  ).
\end{equation}

\begin{equation}
{}^{(3)}R_{nanb} =  {\partial K_{ab} \over \partial n}  + (K^2)_{ab}.
\end{equation}

\noindent
See \cite[page 514, equation (21.76)]{MTW} and \cite[page
516 equation (21.82)]{MTW}. Here the index $n$ refers to the spatial
direction normal to the two-dimensional throat.

Thus far, these results hold both on the throat and in the region
surrounding the throat: these results hold as long as the Gaussian
normal coordinate system does not break down. (Such breakdown being
driven by the fact that the normal geodesics typically intersect after
a certain distance.) In the interests of notational tractability we
now particularize attention to the throat itself, but shall
subsequently indicate that certain of our results can be extended off
the throat itself into the entire region over which the Gaussian normal
coordinate system holds sway.

Taking suitable contractions, 
{\em and using the extremality condition $\tr(K)=0$},

\begin{eqnarray}
{}^{(3)}R_{ab} &=& 
{{}^{(2)}R\over2} \; g_{ab} 
+  {\partial K_{ab} \over \partial n}  
+ 2 (K^2)_{ab}.
\\
{}^{(3)}R_{na} &=&   -K_{ab}{}^{:b}.
\\
{}^{(3)}R_{nn} &=&  
\tr\left({\partial K \over \partial n}\right)  + \tr(K^2)
\\
&=& {\partial\tr(K)\over\partial n} - \tr(K^2).
\end{eqnarray}

\noindent
So that

\begin{eqnarray}
{}^{(3)}R &=&  
{}^{(2)}R 
+  2 \tr\left({\partial K \over \partial n}\right)  
+ 3\tr(K^2)
\\
&=&  
{}^{(2)}R +  2 {\partial \tr(K) \over \partial n}  - \tr(K^2).
\end{eqnarray}
To effect these contractions, we make use of the decomposition of
the three-space metric in terms of the throat two-metric and the
set of two vectors ${e^{i}_a}$ tangent to the throat and the
three-vector $n^i$ normal to the 2-surface

\begin{equation}
{}^{(2+1)}g^{ij} = e^{i}_a \; e^{j}_b \; g^{ab} + n^i n^j.
\end{equation}
For the three-space Einstein tensor ({\em cf.} \cite[page 552]{MTW})
we see

\begin{eqnarray}
{}^{(3)}G_{ab} &=&
{\partial K_{ab} \over \partial n}  + 2 (K^2)_{ab} - 
g_{ab} {\partial \tr(K) \over \partial n} +
{1\over2} g_{ab} \; \tr(K^2).
\\
{}^{(3)}G_{na} &=&   
-K_{ab}{}^{:b}.
\\
{}^{(3)} G_{nn} &=& 
- {1\over2} {}^{(2)} R  - {1\over2} \tr(K^2).
\end{eqnarray}

\noindent
{\em Aside:}
Note in particular that by the flare-out condition ${}^{(3)}R_{nn}
\leq 0$. This implies that the three-space Ricci tensor ${}^{(3)}R_{ij}$
has at least one negative semi-definite eigenvalue everywhere on
the throat. If we adopt the strong flare-out condition then the
three-space Ricci tensor has at least one negative definite eigenvalue
somewhere on the throat. A similar result for Euclidean wormholes
is quoted in~\cite{Giddings-Strominger} and the present analysis
can of course be carried over to Euclidean signature with appropriate
definitional changes.

This decomposition now allows us to write down the various components
of the space-time Einstein tensor. For example

\begin{eqnarray}
{}^{(3+1)}G_{ab} &=& 
 -  \phi_{|ab} - \phi_{|a} \; \phi_{|b}
+ g_{ab} g^{kl} \left[ \phi_{|kl} +  \phi_{|k} \phi_{|l} \right]
\nonumber\\
&&
+{\partial K_{ab} \over \partial n}  + 2 (K^2)_{ab} - 
g_{ab} {\partial \tr(K) \over \partial n} +
{1\over2} g_{ab} \; \tr(K^2)
\nonumber\\
&=&  
8\pi G \; T_{ab}.
\end{eqnarray}

\noindent
But by the definition of the extrinsic curvature, and using the 
Gauss--Weingarten equations,

\begin{eqnarray}
\phi_{|ab} &=& \phi_{:ab} + K_{ab} \; \phi_{|n}.
\\
\phi_{|na} &=&  K_{a}{}^{b} \; \phi_{:b}.
\end{eqnarray}

\noindent
[See, for example, equations (21.57) and (21.63) of~\cite{MTW}.] Thus 

\begin{equation}
g^{kl} \phi_{|kl} = 
g^{ab} \phi_{:ab} +  (g^{ab}K_{ab}) \; \phi_{|n} 
+ \phi_{|nn}.
\end{equation}

\noindent
But remember that $\tr(K)=0$ at the throat, so

\begin{equation}
g^{kl} \phi_{|kl} = 
g^{ab} \phi_{:ab} + \phi_{|nn}.
\end{equation}
This finally allows us to write

\begin{eqnarray}
{}^{(3+1)}G_{ab} 
&=& 
 -  \phi_{:ab} - \phi_{:a} \; \phi_{:b} - K_{ab} \; \phi_{|n}
\nonumber\\
&&
+ g_{ab} \left[ 
 g^{cd} ( \phi_{:cd} +\phi_{:c} \phi_{:d} )
+ \phi_{|nn} + \phi_{|n} \phi_{|n} 
\right]
\nonumber\\
&&
+ {\partial K_{ab} \over \partial n}  + 2 (K^2)_{ab} - 
g_{ab} {\partial \tr(K) \over \partial n} +
{1\over2} g_{ab} \; \tr(K^2)
\nonumber\\
&=&  
8\pi G \; T_{ab}.
\\
{}^{(3+1)}G_{na} 
&=&   
-  K_a{}^b \phi_{:b} - \phi_{|n} \; \phi_{:a}
-  K_{ab}{}^{:b}  
\nonumber\\
&=& 
8\pi G \; T_{na}.
\\
{}^{(3+1)} G_{nn} &=&  g^{cd} 
\left[ \phi_{:cd} + \phi_{:c} \phi_{:d}  \right]
- {1\over2} {}^{(2)} R  - {1\over2} \tr(K^2) 
\nonumber\\
&=& 
-8\pi G \; \tau.
\\
{}^{(3+1)}G_{\hat t a} &=&   0.
\\
{}^{(3+1)}G_{\hat t n} &=&   0.
\\
{}^{(3+1)}G_{\hat t \hat t} &=&  
{{}^{(2)}R\over2} +  
{\partial \tr(K) \over \partial n}  - 
{1\over2} \tr(K^2)
\nonumber\\
&=&
+8\pi G \; \rho.
\end{eqnarray}

\noindent
Here $\tau$ denotes the {\em tension} perpendicular to the wormhole
throat, it is the natural generalization of the quantity considered
by Morris and Thorne, while $\rho$ is simply the energy density at
the wormhole throat.

The calculation presented above is in its own way simply a matter
of brute force index gymnastics---but we feel that there are times
when explicit expressions of this type are useful.

\subsection{Constraints on the stress-energy tensor}

We can now derive several constraints on the stress-energy.

\subsubsection{---First constraint---}

\begin{equation}
\tau = {1\over16\pi G} 
\left[
{}^{(2)} R  +  \tr(K^2)  -2 g^{cd} (\phi_{:cd} + \phi_{:c} \phi_{:d} )
\right].
\end{equation}

\noindent 
(Unfortunately the signs as given are correct. Otherwise we would have
a lovely lower bound on $\tau$. We will need to be a little tricky
when dealing with the $\phi$ terms.)  The above is the generalization
of the Morris--Thorne result that

\begin{equation}
\tau= {1\over8\pi G r_0^2}
\end{equation}

\noindent
at the throat of the special class of model wormholes they
considered. (With MTW conventions ${}^{(2)} R = 2/r_0^2$ for a
two-sphere.) If you now integrate over the surface of the wormhole

\begin{equation}
\int \sqrt{{}^{(2)}g} \; \tau \; d^2x = {1\over16\pi G} 
\left[
4\pi\chi +  \int \sqrt{{}^{(2)}g} 
\left\{\tr(K^2) -2 g^{cd} \phi_{:c} \phi_{:d}\right\} \;d^2 x 
\right].
\end{equation}

\noindent
Here $\chi$ is the Euler characteristic of the throat, while the
$g^{cd} \phi_{:cd}$ term vanishes by partial integration, since the throat
is a manifold without boundary.

\subsubsection{---Second constraint---}

\begin{equation}
\rho = {1\over16\pi G} 
\left[  
{}^{(2)}R +  2 {\partial \tr(K) \over \partial n} - \tr(K^2)
\right].
\end{equation}

\noindent
The second term is negative semi-definite by the flare-out condition,
while the third term is manifestly negative semi-definite. Thus

\begin{equation}
\rho \leq {1\over16\pi G} \; {}^{(2)}R.
\end{equation}

\noindent
This is the generalization of the Morris--Thorne result that 

\begin{equation}
\rho= {b'(r_0)\over8\pi G r_0^2} \leq  {1\over8\pi G r_0^2}
\end{equation}
 
\noindent
at the throat of the special class of model wormholes they considered.
(See~\cite[page 107]{Visser}.)

Note in particular that if the wormhole throat does not have the
topology of a sphere or torus then there {\em must} be places on the
throat such that ${}^{(2)}R < 0$ and thus such that $\rho < 0$.  Thus
wormhole throats of high genus will always have regions that violate
the weak and dominant energy conditions. (The simple flare-out
condition is sufficient for this result. For a general discussion of
the energy conditions see~\cite{Visser} or~\cite{Hawking-Ellis}.)

If the wormhole throat has the topology of a torus then it will
generically violate the weak and dominant energy conditions; only
for the very special case ${}^{(2)}R = 0$, $K_{ab}=0$, $\partial
\tr(K) /\partial n =0$ will it possibly satisfy (but still be on the
verge of violating) the weak and dominant energy conditions. This is
a particular example of a degenerate throat in the sense discussed
previously.

Wormhole throats with the topology of a sphere will, provided they
are convex, at least have positive energy density, but we shall
soon see that other energy conditions are typically violated.

If we now integrate over the surface of the wormhole

\begin{equation}
\int \sqrt{{}^{(2)}g} \; \rho \; d^2x = {1\over16\pi G} 
\left[
4 \pi\chi +  
\int \sqrt{{}^{(2)}g} 
\left\{2 {\partial \tr(K) \over \partial n} - \tr(K^2)\right\} \; d^2 x 
\right].
\end{equation}

\noindent
So for a throat with the topology of a torus ($\chi = 0$) 
the simple flare-out
condition yields

\begin{equation}
\int \sqrt{{}^{(2)}g} \; \rho \; d^2x \leq 0,
\end{equation}

\noindent
while the strong or weak flare-out conditions yield

\begin{equation}
\int \sqrt{{}^{(2)}g} \; \rho \; d^2x < 0,
\end{equation}

\noindent
guaranteeing violation of the weak and dominant energy conditions.
For a throat with higher genus topology $(\chi = 2 - 2g)$ 
the simple flare-out condition
is sufficient to yield

\begin{equation}
\int \sqrt{{}^{(2)}g} \; \rho \; d^2x \leq {\chi\over 4 G} < 0 . 
\end{equation}

\subsubsection{---Third constraint---}

\begin{equation}
\rho-\tau = {1\over16\pi G} 
\left[   
+2 {\partial \tr(K) \over \partial n}
-2 \tr(K^2) 
+2 g^{cd}(\phi_{:cd}  + \phi_{:c} \phi_{:d} )
\right].
\end{equation}

\noindent
Note that the two-curvature ${}^{(2)}R$ has conveniently dropped out
of this equation.  As given, this result is valid only on the
throat itself, but we shall soon see that a generalization can
be constructed that will also hold in the region surrounding the
throat. The first term is negative semi-definite by the simple
flare-out condition (at the very worst when integrated over the throat
it is negative by the weak flare-out condition). The second term is
negative semi-definite by inspection. The third term integrates to
zero though it may have either sign locally on the throat. The fourth
term is unfortunately positive semi-definite on the throat which
prevents us from deriving a truly general energy condition violation
theorem without additional information.

Now because the throat is by definition a compact two surface, we know
that $\phi(x^a)$ must have a maximum somewhere on the throat. At the
global maximum (or even at any local maximum) we have $\phi_{:a}=0$
and $g^{ab} \phi_{:ab}\leq 0$, so at the maxima of $\phi$ one has

\begin{equation}
\rho-\tau \leq 0.
\end{equation}
Generically, the inequality will be strict, and generically there
will be points on the throat at which the null energy condition is
violated.

Integrating over the throat we have

\begin{eqnarray}
&&\int \sqrt{{}^{(2)}g} \; [\rho-\tau] \; d^2 x 
\nonumber\\
&& \qquad
= {1\over16\pi G} 
\int \sqrt{{}^{(2)}g} \left[   
+2 {\partial \tr(K) \over \partial n}
-2 \tr(K^2) 
+2 g^{cd}( \phi_{:c} \phi_{:d} )
\right] d^2x.
\end{eqnarray}

Because of the last term we must be satisfied with the result

\begin{equation}
\int \sqrt{{}^{(2)}g} \; [\rho-\tau] \; d^2 x \leq 
\int \sqrt{{}^{(2)}g} \left[   
2 g^{cd}( \phi_{:c} \phi_{:d} )
\right] d^2x.
\end{equation}

\subsubsection{---Fourth constraint---}

We can rewrite the difference $\rho-\tau$ as

\begin{equation}
\rho-\tau = {1\over16\pi G} 
\left[   
+2 {\partial \tr(K) \over \partial n}
-2 \tr(K^2) 
+2 \exp(-\phi) \;\; {}^{(2)}\Delta \exp(+\phi)
\right].
\end{equation}

\noindent
So if we multiply by $\exp(+\phi)$ before integrating, the
two-dimensional Laplacian  ${}^{(2)}\Delta$
vanishes by partial integration and we have

\begin{eqnarray}
\label{E-t-anec}
&&\int \sqrt{{}^{(2)}g} \; \exp(+\phi) \; [\rho-\tau] \; d^2 x 
\nonumber\\
&& \qquad 
= {1\over8\pi G} 
\int \sqrt{{}^{(2)}g} \; \exp(+\phi) \; 
\left[   
+{\partial \tr(K) \over \partial n}
-\tr(K^2) 
\right] d^2x.
\end{eqnarray}
Thus the strong flare-out condition (or less restrictively, the
weak $e^\phi$--weighted flare-out condition) implies the violation
of this ``transverse averaged null energy condition'' (TANEC, the
NEC averaged over the throat)

\begin{equation}
\int \sqrt{{}^{(2)}g} \; \exp(+\phi) \; [\rho-\tau] \; d^2 x < 0.
\end{equation}
This TANEC, and its off-throat generalization to be developed below,
is perhaps the central result of this generic static wormhole
analysis.

\subsubsection{---Fifth constraint---}

We can define an average transverse pressure on the throat by

\begin{eqnarray}
\bar p &\equiv& {1\over16\pi G} \; \; g^{ab} \;\; {}^{(3+1)}G_{ab}
\\
&=&
{1\over16\pi G}  \left[
g^{cd} ( \phi_{:cd} +\phi_{:c} \phi_{:d} )
+ 2 \phi_{|nn} + 2 \phi_{|n} \phi_{|n} 
- {\partial \tr(K) \over \partial n} + \tr(K^2)
\right].
\nonumber\\
&&
\end{eqnarray}

\noindent
The last term is manifestly positive semi-definite, the penultimate
term is positive semi-definite by the flare-out condition. The
first and third terms are of indefinite sign while the second and
fourth are also positive semi-definite. Integrating over the surface
of the throat

\begin{eqnarray}
\int \sqrt{{}^{(2)}g} \; \bar p  \; d^2x &\geq&
{1\over8\pi G}
\int \sqrt{{}^{(2)}g}  \; \phi_{|nn} \; d^2 x.
\end{eqnarray}
A slightly different constraint, also derivable from the above, is

\begin{eqnarray}
\int \sqrt{{}^{(2)}g} \; e^\phi \; \bar p  \; d^2x &\geq&
{1\over8\pi G}
\int \sqrt{{}^{(2)}g}  \; (e^\phi)_{|nn} \; d^2 x.
\end{eqnarray}
These inequalities relate transverse pressures to normal derivatives of
the gravitational potential. In particular, if the throat lies at a
minimum of the gravitational red-shift the second normal derivative
will be positive, so the transverse pressure (averaged over the
wormhole throat) must be positive.

\subsubsection{---Sixth constraint---}

Now look at the quantities $\rho-\tau+2\bar p$ and $\rho-\tau-2\bar
p$. We have

\begin{eqnarray}
\rho-\tau+2\bar p
&=& 
{1\over4\pi G} \left\{
g^{cd} ( \phi_{:cd} +\phi_{:c} \phi_{:d} )
+  \phi_{|nn} +  \phi_{|n} \phi_{|n} 
\right\}
\\
&=& 
{1\over4\pi G} \left\{
g^{ij} ( \phi_{|ij} +\phi_{|i} \phi_{|j} )
\right\}.
\end{eqnarray}
This serves as a nice consistency check. The combination of
stress-energy components appearing above is equal to $\rho + g^{ij}
T_{ij}$ and is exactly that relevant to the strong energy
condition. See equations
(\ref{E-static-SEC-b})---(\ref{E-static-SEC-e}).  See also equations
(\ref{E-static-stress-energy-b})---(\ref{E-static-stress-energy-e}).
Multiplying by $e^\phi$ and integrating

\begin{eqnarray}
\int \sqrt{{}^{(2)}g} \; e^\phi \; \left[\rho-\tau+2\bar p\right]  \; d^2x =
{1\over4\pi G} \int \sqrt{{}^{(2)}g}  \; (e^\phi)_{|nn} \; d^2 x.
\end{eqnarray}
This relates this transverse integrated version of the strong energy
condition to the normal derivatives of the gravitational potential.

On the other hand

\begin{equation}
\rho-\tau-2\bar p
=
{1\over4\pi G} \left\{
- \phi_{|nn} -  \phi_{|n} \phi_{|n} 
+ {\partial\tr(K)\over\partial n} - \tr(K^2)
\right\}.
\end{equation}
The second and fourth terms are negative semi-definite, while the
third term is negative semi-definite by the flare-out condition.

\subsubsection{---Summary---}

There are a number of powerful constraints that can be placed on the
stress-energy tensor at the wormhole throat simply by invoking the
minimality properties of the wormhole throat. Depending on the precise
form of the assumed flare-out condition, these constraints give the
various energy condition violation theorems we are seeking. Even under
the weakest assumptions (appropriate to a degenerate throat) they
constrain the stress-energy to at best be on the verge of violating
the various energy conditions.

\subsection{Special case: The isopotential throat}

Suppose we take $\phi_{:a}=0$. This additional constraint corresponds
to asserting that the throat is an {\em isopotential} of the
gravitational red-shift. In other words, $\phi(n,x^a)$ is simply
a constant on the throat.  For instance, all the Morris--Thorne
model wormholes~\cite{Morris-Thorne} possess this symmetry. Under
this assumption there are numerous simplifications.

We will not present anew all the results for the Riemann curvature
tensor but instead content ourselves with the Einstein tensor

\begin{eqnarray}
{}^{(3+1)}G_{ab} &=& 
+ g_{ab} \left( 
\phi_{|nn}  + \phi_{|n} \phi_{|n} 
\right) - K_{ab} \; \phi_{|n}
\nonumber\\
&&  
+ {\partial K_{ab} \over \partial n}  + 2 (K^2)_{ab} - 
g_{ab} {\partial \tr(K) \over \partial n} +
{1\over2} g_{ab} \; \tr(K^2)
\nonumber\\
&=&  
8\pi G \; T_{ab}.
\\
{}^{(3+1)}G_{na} &=&   
-K_{ab}{}^{:b}  = 
8\pi G \; T_{na}.
\\
{}^{(3+1)} G_{nn} &=&  
- {1\over2} {}^{(2)} R  - {1\over2} \tr(K^2) = 
-8\pi G \; \tau.
\\
{}^{(3+1)}G_{\hat t a} &=&   0.
\\
{}^{(3+1)}G_{\hat t n} &=&   0.
\\
{}^{(3+1)}G_{\hat t \hat t} &=&  
{{}^{(2)}R\over2} +  
{\partial \tr(K) \over \partial n}- {1\over2} \tr(K^2)
= +8\pi G \; \rho.
\end{eqnarray}

\noindent
Thus for an isopotential throat

\begin{eqnarray}
\tau &=& {1\over16\pi G} 
\left[
{}^{(2)} R  +  \tr(K^2)  
\right] 
\geq {1\over16\pi G} {}^{(2)} R.
\\
\rho &=& {1\over16\pi G} 
\left[  
{}^{(2)}R +  2 {\partial \tr(K) \over \partial n} - \tr(K^2)
\right]  \leq {1\over16\pi G} {}^{(2)} R.
\\
\rho-\tau &=& {1\over16\pi G} 
\left[   
+2 {\partial \tr(K) \over \partial n}
-2 \tr(K^2) 
\right] \leq 0.
\end{eqnarray}

\noindent
This  gives us a very powerful result: using only the simple
flare-out condition, the NEC is on the verge of being violated
everywhere on an isopotential throat.

By invoking the strong flare-out condition the  NEC is definitely
violated somewhere on an isopotential throat.

Invoking the weak flare-out condition we can still say that the
surface integrated NEC is definitely  violated on an isopotential
throat.

\subsection{Special case: The extrinsically flat throat}

Suppose now that we take $K_{ab}=0$. This is a much stronger
constraint than simple minimality of the area of the wormhole throat
and corresponds to asserting that the three-geometry of the throat
is (at least locally) symmetric under interchange of the two regions
it connects.  For instance, all the Morris--Thorne model
wormholes~\cite{Morris-Thorne} possess this symmetry and have
throats that are extrinsically flat.  Under this assumption there
are also massive simplifications. (Note that we are not making the
isopotential assumption at this stage.)

Again, we will not present all the results but content ourselves
with the Einstein tensor

\begin{eqnarray}
{}^{(3+1)}G_{ab} &=& 
 -  \phi_{:ab} - \phi_{:a} \; \phi_{:b}
+ g_{ab} \left[ 
 g^{cd} ( \phi_{:cd} +\phi_{:c} \phi_{:d} )
+ \phi_{|nn}  + \phi_{|n} \phi_{|n} 
\right]
\nonumber\\
&&+ {\partial K_{ab} \over \partial n}  - 
g_{ab} {\partial \tr(K) \over \partial n} 
\nonumber\\
&=&  
8\pi G \; T_{ab}.
\\
{}^{(3+1)}G_{na} &=&    
- \phi_{|n} \; \phi_{|a}  = 
8\pi G \; T_{na}.
\\
{}^{(3+1)} G_{nn} &=&  g^{cd} 
\left[ \phi_{:cd} + \phi_{:c} \phi_{:d}  \right]
- {1\over2} {}^{(2)} R  = 
-8\pi G \; \tau.
\\
{}^{(3+1)}G_{\hat t a} &=&   0.
\\
{}^{(3+1)}G_{\hat t n} &=&   0.
\\
{}^{(3+1)}G_{\hat t \hat t} &=&  
{{}^{(2)}R\over2} +  {\partial \tr(K) \over \partial n}
= +8\pi G \; \rho.
\end{eqnarray}
Though the stress-energy tensor is now somewhat simpler than the
general case, the presence of the $\phi_{:a}$ terms precludes the
derivation of any truly new general theorems.

\subsection{Special case: The extrinsically flat isopotential throat}

Finally, suppose we take both $K_{ab}=0$ and $\phi_{:a}=0$.  A
wormhole throat that is both extrinsically flat and isopotential is
particularly simple to deal with, even though it is still much more
general than the Morris--Thorne wormhole.  Once again, we will not
present all the results but content ourselves with the Einstein tensor

\begin{eqnarray}
{}^{(3+1)}G_{ab} &=& 
+ g_{ab} \left( 
\phi_{|nn} + \phi_{|n} \phi_{|n} 
\right)
+ {\partial K_{ab} \over \partial n}  - 
g_{ab} {\partial \tr(K) \over \partial n}
\nonumber\\
&=&  
8\pi G \; T_{ab}.
\\
{}^{(3+1)}G_{na} &=&  0. 
\\
{}^{(3+1)} G_{nn} &=&  
- {1\over2} {}^{(2)} R  = 
-8\pi G \; \tau.
\\
{}^{(3+1)}G_{\hat t a} &=&   0.
\\
{}^{(3+1)}G_{\hat t n} &=&   0.
\\
{}^{(3+1)}G_{\hat t \hat t} &=&  
{{}^{(2)}R\over2} +  
{\partial \tr(K) \over \partial n}
= +8\pi G \; \rho.
\end{eqnarray}

\noindent
In this case $\rho-\tau$ is particularly simple:

\begin{equation}
\rho - \tau = {1\over8\pi G} {\partial \tr(K) \over \partial n}.
\end{equation}

\noindent
This quantity  is manifestly negative semi-definite by the simple
flare-out condition.
\begin{itemize}
\item
For the strong flare-out condition we deduce that the NEC must be
violated somewhere on the wormhole throat.
\item
Even for the weak flare-out condition we have
\begin{equation}
\int \sqrt{{}^{(2)}g} \; \left[ \rho - \tau \right] \; d^2 x < 0.
\end{equation}
\item
We again see that generic violations of the null energy condition
are the rule.
\end{itemize}

\subsection{The region surrounding the throat}

Because the spacetime is static, one can unambiguously define the
energy density everywhere in the spacetime by setting

\begin{equation}
\rho = { {}^{(3+1)} G_{\hat t\hat t}\over8\pi G}.
\end{equation}
The normal tension, which we have so far defined only on the wormhole
throat itself, can meaningfully be extended to the entire region where
the Gaussian normal coordinate system is well defined by setting

\begin{equation}
\tau = - { {}^{(3+1)} G_{nn}\over8\pi G}.
\end{equation}
Thus in particular

\begin{equation}
\rho-\tau 
= { {}^{(3+1)} G_{\hat t\hat t} + {}^{(3+1)} G_{nn}\over8\pi G} 
= { {}^{(3+1)} R_{\hat t\hat t} + {}^{(3+1)} R_{nn}\over8\pi G},
\end{equation}

\noindent
with this quantity being well defined throughout the Gaussian normal
coordinate patch. (The last equality uses the fact that $g_{\hat
t\hat t}=-1$ while $g_{nn}=+1$.)  But we have already seen how to
evaluate these components of the Ricci tensor. Indeed

\begin{eqnarray}
{}^{(3+1)}R_{\hat t\hat t} &=&  g^{ij}
\left[  
\phi_{|ij} + \phi_{|i} \phi_{|j}
\right].
\\
{}^{(3+1)}R_{nn} &=&  {}^{(3)}R_{nn} -
\left[  
\phi_{|nn} + \phi_{|n} \phi_{|n}
\right]
\\
&=& {\partial\tr(K)\over\partial n} - \tr(K^2) -
\left[  
\phi_{|nn} + \phi_{|n} \phi_{|n}
\right],
\end{eqnarray}

\noindent
where we have been careful to {\em not} use the extremality condition
$\tr(K)=0$. Therefore

\begin{eqnarray}
\rho-\tau
&=& 
{1\over8\pi G}  
\left[ 
{\partial\tr(K)\over\partial n} - \tr(K^2) + 
g^{ab} 
\left(  
\phi_{|ab} + \phi_{|a} \phi_{|b}
\right)
\right]
\\
&=& 
{1\over8\pi G}  
\left[ 
{\partial\tr(K)\over\partial n} - \tr(K^2) + \tr(K) \phi_{|n} +
g^{ab} 
\left(  
\phi_{:ab} + \phi_{:a} \phi_{:b}
\right)
\right],
\nonumber\\
&&
\end{eqnarray}

\noindent
where in the last line we have used the Gauss--Weingarten equations.

\begin{itemize}
\item
If the throat is {\em isopotential}, where isopotential now means
that near the throat the surfaces of constant gravitational potential
coincide with the surfaces of fixed $n$, this simplifies to:
\begin{eqnarray}
\rho-\tau
&=& 
{1\over8\pi G}  
\left[ 
{\partial\tr(K)\over\partial n} - \tr(K^2) + \tr(K) \phi_{|n}
\right].
\end{eqnarray}
\begin{itemize}
\item
If the throat is non-degenerate and satisfies the simple
flare-out condition, then at the throat the first and second terms
are negative semi-definite,  and the third is zero. Then the null
energy condition is either violated or on the verge of being violated
at the throat. 
\item
If the throat is non-degenerate and satisfies the strong
flare-out condition at the point $x$, then  the first term is
negative definite, the second is negative semi-definite, and the
third is zero. Then the null energy condition is violated at the
point $x$ on the throat.  
\item
If the throat satisfies the $N$-fold degenerate flare-out
condition at the point $x$, then by the generalization of the
flare-out conditions applied to degenerate throats the first term
will be $O[n^{2N-2}]$ and negative definite in some region surrounding
the throat. The second term is again negative semi-definite.  The
third term can have either sign but will be $O[n^{2N-1}]$. Thus
there will be some region $n\in(0,n_*)$ in which the first term
dominates.  Therefore the  null energy condition is violated along
the line $\{x\}\times(0,n_*)$. If at every point $x$ on the throat
the $N$-fold degenerate flare-out condition is satisfied for some
{\em finite} $N$, then there will be an open region surrounding
the throat on which the null energy condition is everywhere violated.
\item
This is the closest one can get in generalizing to arbitrary
wormhole shapes the discussion on page 405 [equation (56)] of
Morris--Thorne~\cite{Morris-Thorne}.  Note carefully their use of
the phrase  ``at or near the throat''. In our parlance, they are
considering a spherically symmetric extrinsically flat isopotential
throat that satisfies the $N$-fold degenerate flare-out condition
for some finite but unspecified $N$. See also page 104, equation
(11.12) and page 109, equation (11.54) of~\cite{Visser}, and
contrast this with equation (11.56).
\end{itemize}
\item
If the throat is not isopotential we multiply by $\exp(\phi)$ and
integrate over surfaces of constant $n$. Then
\begin{eqnarray}
&&\int \sqrt{{}^{(2)} g} \; \exp(\phi) \; [\rho-\tau] \;  d^2x =
\nonumber\\
&&\qquad
{1\over8\pi G}  \int \sqrt{{}^{(2)} g} \; \exp(\phi)
\left[ 
{\partial\tr(K)\over\partial n} - \tr(K^2) + \tr(K) \phi_{|n}
\right]
 d^2x.
\nonumber\\
\end{eqnarray}
This generalizes the previous version (\ref{E-t-anec}) of the
transverse averaged null energy condition to constant $n$ hypersurfaces
near the throat.  For each point $x$ on the throat, assuming the
$N$-fold degenerate flare-out condition,  we can by the previous
argument find a range of values [$n\in(0,n_*(x))$] that will make
the integrand negative. Thus there will be a set of values of $n$
for which the integral is negative.  Again we deduce violations of
the null energy condition.
\end{itemize}

\section{Discussion}

In this survey we have sought to give an overview of the energy
condition violations that occur in traversable wormholes. We point
out that in static spherically symmetric geometries these violations
of the energy conditions follow unavoidably from the definition of
a wormhole and the definition of the total stress-energy via the
Einstein equations. In spherically symmetric time dependent situations
limited temporary suspensions of the energy condition violations
are possible. In non-Einstein theories of gravity it is often
possible to push the energy condition violations into the nonstandard
parts of the stress-energy tensor and let the ordinary part of the
stress-energy satisfy the energy conditions. The total stress-energy
tensor, however, must still violate the energy conditions. 

To show the generality of the energy condition violations, we have
developed an analysis that is capable of dealing with static
traversable wormholes of arbitrary shape.  We have presented a
definition of a wormhole throat that is much more general than that
of the Morris--Thorne wormhole~\cite{Morris-Thorne}.  The present
definition works well in any static spacetime and nicely captures
the essence of the idea of what we would want to call a wormhole
throat.

We do not need to make any assumptions about the existence of any
asymptotically flat region, nor do we need to assume that the
manifold is topologically non-trivial. It is important to realise
that the essence of the definition lies in the geometrical structure
of the wormhole throat.

Starting from our definition we have used the theory of embedded
hypersurfaces to place restrictions on the Riemann tensor and
stress-energy tensor at the throat of the wormhole. We find, as
expected, that the wormhole throat generically violates the null
energy condition and we have provided several theorems regarding
this matter. These theorems generalize the Morris--Thorne results
on exotic matter~\cite{Morris-Thorne}, and are complementary to
the topological censorship theorem~\cite{Topological-censorship}.

Generalization to the time dependent situation is clearly of
interest. Unfortunately we have encountered many subtleties of
definition, notation, and formalism in this endeavor. (A formulation
in terms of {\em anti-trapped surfaces} appears promising~\cite{Page}.)
We defer the issue of time dependent wormhole throats to a future
publication.

\section*{Acknowledgements}

M.V. wishes to gratefully acknowledge the hospitality shown during
his visits to the Laboratory for Space Astrophysics and Fundamental
Physics (LAEFF, Madrid). This work was supported in part by the US
Department of Energy (M.V.) and by the Spanish Ministry of Science
and Education (D.H.).




\begin{thebibliography}{99}

\bibitem{Morris-Thorne} 
M.S. Morris and K.S. Thorne, 
Am. J. Phys. {\bf 56}, 395 (1988).

\bibitem{MTY} 
M.S. Morris, K.S. Thorne and U. Yurtsever, 
Phys. Rev. Lett, {\bf 61}, 1446 (1988).

\bibitem{Visser} 
M. Visser, 
{\em Lorentzian Wormholes: From Einstein to Hawking}
(American Institute of Physics, Woodbury, N.Y., 1995).

\bibitem{HPS} 
D. Hochberg, A. Popov and S.V. Sushkov, 
Phys. Rev. Lett. {\bf 78}, 2050 (1997).

\bibitem{Topological-censorship} 
J.L Friedmann, K. Schleich and D.M. Witt,
Phys. Rev. Lett. {\bf 71}, 1486 (1993).

\bibitem{Page}
D. Page, cited as a note added in proof in~\cite{Morris-Thorne}.

\bibitem{Frolov-Novikov:device}
V. Frolov and I. D.  Novikov
Phys. Rev. D48 (1993) 1607-1615.

\bibitem{Wheeler:56}
J. A. Wheeler,
Phys. Rev. 97 (1955) 511-536.

\bibitem{Wheeler:57}
J. A. Wheeler, 
Ann. Phys. (NY) 2 (1957) 604--616.


\bibitem{Hawking:cpc}
S. W. Hawking,
Phys. Rev. D46 (1992) 603--611.

\bibitem{Visser:reliability}
M. Visser,  
{\em The reliability horizon for semiclassical quantum gravity:
Metric fluctuations are often more important than back reaction.}
gr-qc/9702041; Physics Letters B, in press.

\bibitem{gvp:anec}
M. Visser,
gr-qc/9409043, Phys. Lett. B349 (1995) 443--447.	   

\bibitem{gvp1}
M. Visser,
gr-qc/9604007, Phys. Rev. D54 (1996) 5103--5115.

\bibitem{gvp2}
M. Visser
gr-qc/9604008, Phys. Rev. D54 (1996) 5116--5122.

\bibitem{gvp3}
M. Visser,
gr-qc/9604009, Phys. Rev. D54 (1996) 5123--5128.

\bibitem{gvp4}
M. Visser,
gr-qc/9703001, Phys. Rev. D56 (1997) 936--952.

\bibitem{Visser:quantum-wormholes}
M. Visser,
Phys. Rev. D43 (1991) 402--409.

\bibitem{Visser:wormshop}
M. Visser,
Phys. Lett. B242 (1990) 24--28.

\bibitem{Visser:rice}
M. Visser, {\em Quantum wormholes in Lorentzian signature}, in B.
Bonnor and H. Miettinen, editors, {\em Proceedings of the Rice
meeting: 1990 meeting of the Division of Particles and Fields of
the American Physical Society}, volume 2, pages 858--860. (World
Scientific, Singapore, 1990).

\bibitem{Aichelburg}
F. Schein and P.C. Aichelburg, 
gr-qc/9606069, Phys. Rev. Lett. 77 (1996) 4130-4133.

\bibitem{Aichelburg2}
F. Schein, P.C. Aichelburg, and W. Israel,
gr-qc/9602053, Phys. Rev. D54 (1996) 3800-3805.

\bibitem{Visser89a} 
M. Visser, Phys. Rev. {\bf D}39, 3182 (1989).

\bibitem{Roman}
T. A. Roman,
Phys. Rev. D47 (1993) 1370--1379.

\bibitem{Hochberg-Kephart}
D. Hochberg and T. W. Kephart,
Phys. Rev. Lett 70 (1993) 2665--2668.

\bibitem{Kar}
S. Kar, 
Phys. Rev. D49 (1994) 862--865.

\bibitem{Kar2}
S. Kar and D. Sahdev,
Phys. Rev. D53 (1996) 722--730.

\bibitem{Kim}
S. W. Kim,
Phys. Rev. D53 (1996) 6889--6892.

\bibitem{Agnese}
A. G. Agnese and M. La Camera,
Phys. Rev. D51 (1995) 2011--2013.

\bibitem{Nandi}
K. K. Nandi, A. Islam, and J. Evans,
Phys. Rev. D55 (1997) 2497--2450.

\bibitem{Anchordoqui}
L. A. Anchordoqui, S. Perez Bergliaffa, and D. F. Torres,
Phys. Rev. D55 (1997) 5226--5229.

\bibitem{Yang-Zhang}
H. H. Yang and Y. Z. Zhang,
Phys. Lett. A 212 (1996) 39--42.

\bibitem{Hochberg-Visser}
D. Hochberg and M. Visser, 
{\em Geometric structure of the generic static traversable wormhole throat},
gr-qc/9704082; Phys. Rev. D (in press).

\bibitem{Israel67}
W. Israel, 
Phys. Rev. 164 (1967) 1776--1779.

\bibitem{Israel68}
W. Israel, 
Commun. Math. Phys. 8 (1968) 254--260.

\bibitem{Israel86}
W. Israel, 
Can. J. Phys. 64 (1986) 120--127.

\bibitem{MTW} 
C.W. Misner, K.S. Thorne and J.A. Wheeler 
{\em Gravitation} 
(W.H. Freeman, San Francisco, 1973).


\bibitem{Hawking-Ellis} 
S.W. Hawking and G.F.R. Ellis,
{\em The Large Scale Structure of Space-Time}
(Cambridge University Press, Cambridge, England, 1973).

\bibitem{Wald} 
R.M. Wald, 
{\em General Relativity} 
(University of Chicago Press, Chicago, 1984).

\bibitem{Synge} 
J.L. Synge, {\em Relativity: the General Theory}
(North-Holland, Amsterdam, 1964).

\bibitem{Fock} V. Fock, 
{\em The Theory of Space, Time, and Gravitation}
(Pergamon, New York, 1964).

\bibitem{Adler-Bazin-Schiffer} 
R. Adler, M. Bazin, and M. Schiffer,
{\em Introduction to General Relativity}
(McGraw--Hill, New York, 1965).


\bibitem{Visser89b} 
M. Visser, Nucl. Phys. {\bf B}328, 203 (1989).

\bibitem{Giddings-Strominger}
S. Giddings and A. Strominger,
Nucl. Phys. B306 (1988) 890--907, see esp.~page 894.


\end{thebibliography}
\end{document}